\documentclass[prc,showpacs,floatfix]{revtex4}
\usepackage{bm}
\usepackage{dcolumn}
\usepackage{graphicx}
\usepackage{amsmath}

\newcommand{\beq}{\begin{equation}}
\newcommand{\eeq}{\end{equation}}
\newcommand{\beqa}{\begin{eqnarray}}
\newcommand{\eeqa}{\end{eqnarray}}

\begin{document}
\title{
\hfill{\small {\bf MKPH-T-10-19}}\\
{\bf Polarization observables in $\pi^0\eta$-photoproduction on the proton}}
\author{A. Fix and H. Arenh\"ovel }
\affiliation{
Institut f\"ur Kernphysik,
Johannes Gutenberg-Universit\"at Mainz, D-55099 Mainz, Germany}
\date{\today}
\begin{abstract}
For $\pi^0\eta$-photoproduction on the nucleon  formal expressions are
developed for the five-fold differential cross section and the
recoil polarization including beam and target
polarizations. The polarization observables are described by various beam, target and
beam-target asymmetries for polarized photons and/or polarized nucleons. They are given as
bilinear hermitean forms in the reaction matrix elements divided by the unpolarized cross
section. Numerical results for the linear and circular beam asymmetries for $\gamma
p\to\pi^0\eta p$ are obtained within an isobar model and are compared with existing data.
Predictions are also given for the target asymmetry $T_{11}$, and the
beam-target asymmetries $T^c_{10}$ and $T^c_{11}$ for circularly
polarized photons.
\end{abstract}

\pacs{13.60.Le, 13.75.-n, 21.45.+v, 25.20.Lj}
\maketitle

\section{Introduction}

Polarization observables are known to be an essential ingredient in the interpretation of
photon induced meson production reactions, especially if the
production process proceeds predominantly via
resonance excitations. Their study provides further insight into the details of the underlying
reaction mechanisms and possible structure effects. Thus, such observables will
serve as an additional critical test for theoretical models.

Today special interest is focussed to processes with more than a
single pseudoscalar meson in the final state. These reactions constitute
a rather new object in particle physics. At present most of the
efforts are directed towards an understanding of their general dynamical
content. In such a situation experiments with polarized particles are
therefore of special use.
Different analyses clearly demonstrate their importance, primarily since the unpolarized
data are usually unable to impose sufficient constraints on the model parameters.

Experiments for $\pi\pi$ and $\pi^0\eta$ photoproduction have become a
center of attention in recent research programs discussed at ELSA
and MAMI~\cite{Weinheimer,Horn,Tohoku,Ajaka,Kashev} . A major point of
these programs is a study 
of those resonances for which only a weak evidence exists. It is therefore timely to
investigate in detail the polarization structure of double meson photoproduction. Some
important steps towards this goal are already done in Ref.~\cite{Roberts}, where in
particular a set of polarization experiments, needed to determine the
reaction amplitude, is discussed.

With the present work we want to provide a complete solid basis for the formal expressions of all
possible polarization observables which determine the general differential cross section and the
proton recoil polarization for $\pi^0\eta$ production on a polarized proton target with
polarized photons in a compact and suggestive notation.

Our second goal is to study the properties of those observables for which experimental
results already exist or are expected to be measured in the near future. Recently, polarization
measurements of different beam asymmetries in $\pi^0\eta$ photoproduction were performed
at ELSA~\cite{GutzBeam,Gutz_Is} for the first time. Furthermore, new
MAMI results for the target asymmetry $T^0_{11}$
and the beam-target asymmetry $T ^c_{11}$  are now expected. Here, we
pay some attention to the properties of the 
circular beam asymmetry measured at MAMI~\cite{KashevAphi}. The analysis of this observable
for the similar reaction, $\pi\pi$
photoproduction~\cite{Strauch,Kram,Roca}, confirms a strong
sensitivity of the data to the dynamical content of the amplitude. As will be shown in
the present paper, the information contained in the circular asymmetry provides
constraints on the contribution of positive parity resonances to
$\pi^0\eta$ photoproduction.

The paper is organized as follows. In the next three sections we develop the basic
formalism for the differential cross section with inclusion of polarization observables.
In Sect.~\ref{results}, the most essential ingredients for the calculation of the $T$
matrix in the isobar model are described. Here we also present and discuss the results on
some beam, target, and beam-target asymmetries, which  are also compared to the
existing data. In several appendices we describe in detail some ingredients of our formal
developments. One should note that throughout this paper $\pi$ meson
always means $\pi^0$ meson.

\section{Kinematics}

\begin{figure}[htb]
\includegraphics[scale=.8]{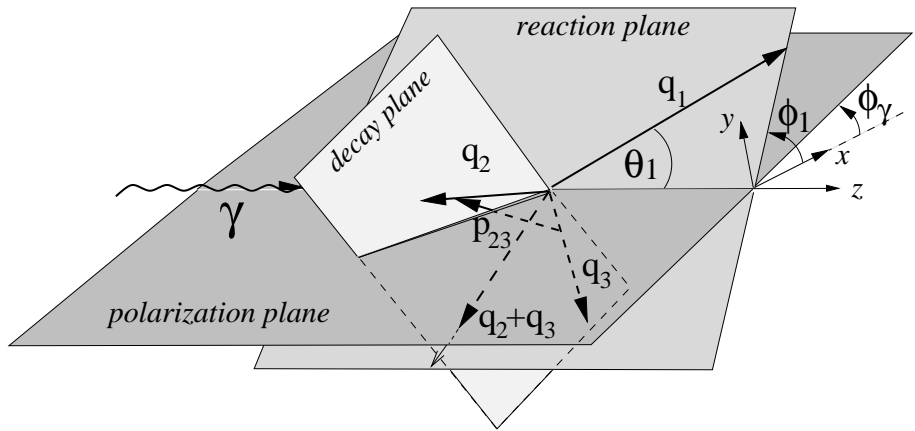}
\caption{Kinematics of $\pi\eta$ photoproduction on the nucleon in the
  laboratory system. The active particle is denoted by ``1'' and
  defines  the reaction plane. The nonrelativistic relative momentum is
  denoted by $\vec p_{23}=(m_3\vec q_2 -m_2\vec q_3)/(m_2+m_3) $.}
\label{fig_kinematics}
\end{figure}

As a starting point, we first will consider the kinematics of the photoproduction
reaction
\begin{equation}
\gamma(k,\vec{\varepsilon}_\mu)+N_i(p_i)\!\rightarrow\! \pi(q_\pi)+\eta(q_\eta)+N_f(p_f),
\end{equation}
defining the notation of the four-momenta of the participating particles
\begin{equation}
k=(\omega_\gamma,\vec k)\,,\quad p_i=(E_i,\vec p_i)\,,\quad q_\pi=(\omega_\pi,\vec
q_\pi)\,,\quad q_\eta=(\omega_\eta,\vec q_\eta)\,,\quad p_f=(E_f,\vec p_f)\,.
\end{equation}
As coordinate system we choose a right-handed one with $z$-axis along
the photon momentum $\vec{k}$ and the other axis perpendicular. As is
illustrated in Fig.~\ref{fig_kinematics} for the laboratory frame, we
distinguish three planes: 
\begin{description}
\item[(i)]
The reaction plane, spanned by the momenta of 
incoming photon $\vec k $ and $\vec q_1$ of particle ``1'', called 
the active particle, which usually is 
detected. This plane intersects the $x$-$z$-plane 
 along the $z$-axis with an angle~$\phi_1$.
\item[(ii)]
The polarization or photon plane, spanned by the photon momentum
and the direction of maximal linear photon polarization, which
intersects the $x$-$z$-plane along the $z$-axis with an angle
$\phi_\gamma$ and the reaction plane along the $z$-axis
with an angle $|\phi_1-\phi_\gamma|$. 
\item[(iii)]
The decay plane, spanned by the momenta of the
other two outgoing particles ``2'' and ``3'',  intersecting the
reaction plane along the total momentum $\vec p_2 + \vec p_3$ of the
latter two particles. 
In case that the linear photon polarization vanishes, one can
choose $\phi_1=\phi_\gamma=0$ and then polarization and reaction planes
coincide. 
\end{description}

The following formal developments will not depend on whether one
chooses as reference frame the laboratory or the center-of-momentum
(c.m.) frame. Furthermore, we will consider the $\eta$ meson as
particle ``1''  ($\vec q_1:=\vec q_\eta$) defining the reaction plane, 
while pion and proton constitute
particles ``2'' and ``3'', respectively, in the decay plane, i.e.\
$\vec q_2:=\vec q_\pi $ and $\vec q_3:=\vec p_f $.
Besides the incoming photon momentum $\vec k$, we choose as
independent variables for the description of cross 
section and polarization observables the angle $\phi_\gamma$
characterizing the polarization plane, the outgoing $\eta$ momentum 
$\vec q_\eta=(q_\eta,\theta_\eta,\phi_\eta)$,
and the spherical
angles $\Omega_{\pi p}=(\theta_{\pi p},\phi_{\pi p})$ of the relative
momentum $\vec p_{\pi p}$ of the outgoing pion and nucleon as given by
\beq
\vec p_{\pi p}=(M_p\vec
q_\pi-m_\pi\vec p_f)/(M_p+m_\pi)=(p_{\pi p},\Omega_{\pi p})\,.\label{momentum-p}
\eeq
Then the momenta of outgoing pion and nucleon are fixed. For example,
the pion momentum reads 
\begin{equation}\label{qp}
\vec q_{\pi}=\vec p_{\pi p}+\frac{m_\pi}{M_p+m_\pi} (\vec k +\vec p_i -{\vec q_\eta})\,.
\end{equation}

In Sect.~\ref{results} we will also consider configurations where
either the outgoing pion or proton is the active particle, i.e.\
constituting the reaction plane, while the decay plane is spanned
by the momenta of the other two particles in the final state, i.e.\
either eta and proton or pion and eta, respectively.

\section{The $T$-matrix}
\label{Tmatrix}
To be specific, we take in this section the outgoing eta as active
particle. The corresponding expressions for the pion as active
particle are simpy obtained by the interchange
$\eta\leftrightarrow\pi$. For the outgoing proton as active particle,
detailed expressions are listed in the Appendix~\ref{appa5}.

All observables are determined by the $T$-matrix elements of the electromagnetic
$\pi\eta$ production current $\vec J_{\gamma\pi\eta}$ between the initial proton and the
final outgoing $\pi\eta N_f$ scattering states (indicated by a
supersript ``(-)''). In a general frame it is given by
\begin{equation}
T_{m_f \mu m_i}= -^{(-)}\langle \vec q_\eta,\,\vec q_\pi;\, \vec p_f\,m_f\,|\
\vec\varepsilon_\mu\cdot\vec J_{\gamma\pi\eta}(0)| \vec p_i\, m_i\rangle\,,
\end{equation}
where $m_f$ denotes the proton spin projection on the relative
momentum $\vec p_{\pi p}$ of the
outgoing pion and proton, and $m_i$ correspondingly the initial proton spin projection
on the $z$-axis as quantization axis. The circular polarization vector
of the photon is denoted by $\vec{\varepsilon}_\mu$ with
$\mu=\pm 1$. Furthermore, transverse gauge has been chosen. The
knowledge of the specific form of $\vec J_{\gamma\pi\eta}$ is not needed for the
following formal considerations.

The general form of the $T$-matrix after separation of the overall
c.m.-motion and insertion of the multipole expansion of the current
operator is given in terms of the relative $\pi p$ momentum and the
$\eta$ momentum by
\begin{eqnarray}
T_{m_f \mu m_i}(\vec p_{\pi p},\vec q_\eta)&=& -
^{(-)}\langle \vec p_{\pi p}\, m_f;\,\vec q_\eta\,|J_{\gamma\pi\eta,\,\mu}(\vec k\,)|m_i\rangle\nonumber\\
&=& \sqrt{2\pi}\sum_{L} i^L\widehat L ^{(-)}\langle \vec p_{\pi p}\, m_f;\,\vec q_\eta\,|{\cal
O}^{\mu L}_\mu|m_i\rangle\,,
\end{eqnarray}
with $\mu=\pm 1$,  $\widehat L=\sqrt{2L+1}$, and transverse multipoles
\begin{eqnarray}
{\cal O}^{\mu L}_M&=& E_M^L +\mu M_M^L\,.
\end{eqnarray}
It is convenient to introduce a partial wave decomposition of the
final outgoing scattering state
\begin{eqnarray}
^{(-)}\langle \vec p_{\pi p}\, m_f|&=&\frac{1}{\sqrt{4\pi}} \sum_{l_{\pi p} j_{\pi p} m_{\pi p}}\widehat l_{\pi p}\, (l_{\pi p}
0 \frac{1}{2} m_f|j_{\pi p} m_f)\, D^{j_{\pi p}}_{m_f,m_{\pi p}}(\phi_{\pi p},-\theta_{\pi p},-\phi_{\pi p})
^{(-)}\langle p_{\pi p} \, (l_{\pi p}  \frac{1}{2})j_{\pi p} m_{\pi p}|\,,\\
^{(-)}\langle \vec q_\eta\,|&=&\frac{1}{\sqrt{4\pi}}
\sum_{l_\eta m_\eta}\widehat l_\eta\,D^{l_\eta}_{0,m_\eta}( \phi_\eta,-\theta_\eta,-\phi_\eta)
^{(-)}\langle q_\eta l_\eta m_\eta|\,,
\end{eqnarray}
where $m_{\pi p}$ and $m_\eta$ like $m_i$ refer to the photon momentum $\vec k$ as
quantization axis. Here, the rotation matrices $D^j_{m'm}$ are taken in the convention of
Rose~\cite{Ros57}.
Using the Wigner-Eckart theorem, one obtains
\begin{eqnarray}
^{(-)}\langle p_{\pi p}\, (l_{\pi p} \frac{1}{2})j_{\pi p} m_{\pi p}; q_\eta\,l_\eta
m_\eta|{\cal O}^{\mu L}_\mu| \frac{1}{2} m_i\rangle &=&\sum_{J M_J}(-1)^{j_{\pi
p}-l_\eta+J}\,\widehat{J} \left(\begin{array}{ccc} j_{\pi p} & l_\eta & J \cr m_{\pi p} &
m_\eta & -M_J\cr
\end{array}\right)
\left(\begin{array}{ccc}
J & L & \frac{1}{2} \cr -M_J & \mu & m_i\cr
\end{array}\right)\nonumber\\&&\hspace*{1.5cm}\times
\langle p_{\pi p}\, q_\eta ;((l_{\pi p} \frac{1}{2})j_{\pi p}
l_\eta)J||{\cal O}^{\mu L}|| \frac{1}{2}\rangle\,, \label{mat1}
\end{eqnarray}
with the selection rule $m_{\pi p}+m_\eta=M_J=\mu+m_i$.
The angular dependence can be rewritten according to
\begin{eqnarray}
D^{j_{\pi p}}_{m_f,m_{\pi p}}(\phi_{\pi p},-\theta_{\pi p},-\phi_{\pi p})\,
D^{l_\eta}_{0,m_\eta}(\phi_\eta,-\theta_\eta,-\phi_\eta)&=& d^{j_{\pi p}}_{m_f,m_{\pi p}}(-\theta_{\pi p})
\,d^{\,l_\eta}_{0,m_\eta}(-\theta_\eta)\, e^{i((m_{\pi p}-m_f)\phi_{\pi p}+m_\eta\phi_\eta)}
\,,
\end{eqnarray}
where $d^j_{m m'}$ denotes a small rotation matrix~\cite{Ros57}.
Rearranging
\begin{eqnarray}
(m_{\pi p}-m_f)\phi_{\pi p}+m_\eta\phi_\eta&=&(m_{\pi p}-m_f)\phi_{pq}+(\mu+m_i-m_f)\phi_\eta
\end{eqnarray}
with $\phi_{pq}=\phi_{\pi p}-\phi_{\eta}$, one finds that the dependence on $\phi_\eta$ can be
separated, i.e.\
\begin{eqnarray}\label{small_t}
T_{m_f \mu m_i}(\Omega_{\pi p},\Omega_\eta)&=& e^{i(\mu+m_i-m_f)\phi_\eta} t_{m_f \mu
m_i}(\theta_{\pi p},\,\theta_\eta,\,\phi_{pq})\,,
\end{eqnarray}
where the small $t$-matrix depends only on $\theta_{\pi p}$, $\theta_\eta$, and the relative
azimuthal angle $\phi_{pq}$.

Explicitly, one obtains
\begin{eqnarray}
t_{m_f \mu m_i}(\theta_{\pi p},\,\theta_\eta,\,\phi_{pq})&=& \frac{1}{2\,\sqrt{2\pi}} \sum_{L
l_{\pi p} j_{\pi p} m_{\pi p} l_\eta m_\eta J M_J}i^L\,\widehat L\,\widehat J\,\widehat l_\eta
\,\widehat l_{\pi p}\,\widehat j_{\pi p}\,(-1)^{J+l_{\pi p}+j_{\pi p}-\frac{1}{2}+m_f-l_\eta}\nonumber\\
&&\times\left(\begin{array}{ccc} l_{\pi p} &  \frac{1}{2} & j_{\pi p} \cr 0 & m_f & -m_f\cr
\end{array}\right)
\left(\begin{array}{ccc} j_{\pi p} & l_\eta & J \cr m_{\pi p} & m_\eta & -M_J\cr
\end{array}\right)
\left(\begin{array}{ccc}
J & L & \frac{1}{2} \cr -M_J & \mu & m_i\cr
\end{array}\right)\nonumber\\
&&\times\langle p_{\pi p} \,q_\eta; ((l_{\pi p} \frac{1}{2})j_{\pi p} l_\eta)J||{\cal O}^{\mu L}||\frac{1}{2}\rangle
d^{j_{\pi p}}_{m_f,m_{\pi p}}(-\theta_{\pi p})\,d^{l_\eta}_{0,m_\eta}(-\theta_\eta)\,
e^{i(m_{\pi p}-m_f)\phi_{p q}}\,.\label{smallt}
\end{eqnarray}

In the case that parity is conserved, it is quite
straightforward to show that, the following
symmetry relation holds for the inverted spin projections of  the small
$t$-matrix elements
\begin{eqnarray}\label{symmetry}
t_{-m_f -\mu -m_i}(\theta_{\pi p},\,\theta_\eta,\,\phi_{pq})&=& (-1)^{-m_f+\mu+m_i}
t_{m_f \mu m_i}(\theta_{\pi p},\,\theta_\eta,\,-\phi_{pq})\,.
\end{eqnarray}
Besides the phase factor, one should note the sign change of $\phi_{p q}$ on the
right-hand side. In the derivation of this relation one has made use of the parity
selection rules for the multipole transitions to a final partial wave $|p q ((l_{\pi
p}s)j_{\pi p} l_\eta)J\rangle$ with parity $\pi_{J(l_{\pi p},l_\eta)}=(-1)^{l_{\pi
    p}+l_\eta}$, which read
\begin{eqnarray}
\begin{array}{lll} E^L: & \pi_i \pi_{J(l_{\pi p},l_\eta)}\,(-1)^L=1 &\rightarrow
(-1)^{l_{\pi p}+l_\eta+L}=1\,,\cr M^L: & \pi_i \pi_{J(l_{\pi p},l_\eta)}\,(-1)^L=-1
&\rightarrow (-1)^{l_{\pi p}+l_\eta+L}=-1\,.\cr
\end{array}
\end{eqnarray}
Therefore, invariance under a parity transformation results in the
following property of the reduced matrix element
\begin{eqnarray}
(-1)^{l_{\pi p}+l_\eta+L}\langle p_{\pi p} \, q_\eta; ((l_{\pi p}\frac{1}{2})j_{\pi p} l_\eta)J||
{\cal O}^{-\mu L}||\frac{1}{2}\rangle&=& \langle p_{\pi p}\, q_\eta; ((l_{\pi p} \frac{1}{2})j_{\pi
p} l_\eta)J||{\cal O}^{\mu L}|| \frac{1}{2}\rangle\,.
\end{eqnarray}
A  corresponding relation for the $T$-matrix elements follows from the symmetry
property (\ref{symmetry})
\begin{eqnarray}
T_{-m_f -\mu -m_i}(\theta_{\pi p},\phi_{\pi p},\theta_\eta,\phi_\eta)&=&
(-1)^{-m_f+\mu+m_i} T_{m_f \mu m_i}(\theta_{\pi p},-\phi_{\pi
p},\theta_\eta,-\phi_\eta)\,.
\end{eqnarray}
These symmetry properties are valid for all three choices of the
active particle. 

Since the small $t$-matrix elements are the basic quantities, which determine
the general differential cross section and the recoil polarization in
terms of bilinear hermitean forms in the $t$-matrix elements, the
developments of the next section are independent of which particle is
chosen as active. 

\section{Differential cross section and recoil polarization}
\label{diff_cross}

The starting point for the formal derivation of polarization observables is the
evaluation of the following trace with respect to the spin degrees of freedom
\begin{equation}
A_{I'M'}=c(q_\eta,\theta_\eta,\Omega_{\pi p})\, tr(T^\dagger
\tau^{f,[I']}_{M'}T\rho_i)\,,\label{trace}
\end{equation}
for $I'=0,1$ and $M'=-I',\dots,I'$, folded between the density matrix $\rho_i$ for
the spin degrees of the initial system and a spin operator $\tau^{f,[I']}_{M'}$ with
respect to the final nucleon spin space. The latter is defined by its reduced matrix
element
\begin{equation}
\langle \frac{1}{2}||\tau^{[I']}||\frac{1}{2} \rangle = \sqrt{2}\,\widehat {I'}
\quad\mbox{for}\quad I'=0,1\,.\label{tau-operator}
\end{equation}
Note that $\tau^{[1]}$ corresponds to the conventional Pauli spin operator $\vec \sigma$.
The trace refers to all initial and final state spin degrees of freedom of incoming
photon, target and recoiling nucleon. The kinematic factor
$c(q_\eta,\theta_\eta,\Omega_{\pi p})$ comprises the final state phase space and the incoming
flux.
In an arbitrary frame one has
\beq
c(q_\eta,\Omega_q,\Omega_{\pi p})=\frac{1}{(2\pi)^5}\,\frac{M_p^2}{E_{i}+p_i}\,\frac{1}{8\omega_\gamma\omega_\eta}
\frac{p_{\pi p}^2}{p_{\pi
    p}(\omega_\pi+E_{f})+\frac{(\vec{q}_\pi+\vec{p}_f)\cdot\vec{p}_{\pi
      p}}{p_{\pi p}(M_p+m_\pi)}(E_{f}m_\pi-\omega_\pi M_p)}\,.
\eeq

The general expression for the differential cross section is given by
\begin{equation}
\frac{d\sigma}{d\vec q_\eta d\Omega_{\pi p}}=A_{00}\,,\label{diffcrossx}
\end{equation}
and the final nucleon polarization component $P_M$ with respect to a spherical basis
\begin{equation}
P_M\frac{d\sigma}{d\vec q_\eta d\Omega_{\pi p}}=(-1)^MA_{1-M}\, .\label{recoilpol}
\end{equation}
With respect to a cartesian basis, one has as polarization components
\begin{equation}
P_x\frac{d\sigma}{d\vec q_\eta d\Omega_{\pi p}}=\frac{1}{\sqrt{2}}\,B^-_1 \,,\quad
P_y\frac{d\sigma}{d\vec q_\eta d\Omega_{\pi p}}=\frac{i}{\sqrt{2}}\,B^+_1 \,,\quad
P_z\frac{d\sigma}{d\vec q_\eta d\Omega_{\pi p}}=B^+_0\,,\label{cartesian-P}
\end{equation}
where for $M=0,1$ we have introduced
\begin{equation}
B^\pm_{M}=\frac{(-1)^M}{1+\delta_{M0}}\,(A_{1M}\pm A_{1-M})\,.\label{B_M}
\end{equation}

The density matrix $\rho_i$ in (\ref{trace}) is a direct product of the density matrices
$\rho^\gamma$ of the photon and $\rho^p$ of the nucleon
\begin{equation}
\rho_i=\rho^\gamma\otimes\rho^p \,.
\end{equation}
For the chosen reference frame, the photon density matrix has the form
\begin{equation}
\rho^\gamma_{\mu \mu'}=\frac{1}{2}(\delta_{\mu \mu'}+ \vec P^\gamma\cdot\vec \sigma_{\mu
\mu'})
\end{equation}
with respect to the circular polarization basis ($\mu=\pm 1$). Here,
$|\vec P^\gamma|$ describes the
total degree of polarization, $P^\gamma_z=P^\gamma_c$ is the
difference of right to left circularly polarized photons, i.e.\
$|P^\gamma_c|$ describes the degree of circular polarization being
right or left according to whether $P^\gamma_c>0$ or $<0$,
respectively, and
$P^\gamma_l=\sqrt{(P^\gamma_x)^2+(P^\gamma_y)^2}$ describes the degree of linear
polarization. By a rotation around the photon momentum by an
appropriate angle $\phi_\gamma$ it is possible to have the new
$x'$-axis pointing in the direction of maximum linear polarization.
Then one has $P^\gamma_{x'}=-P^\gamma_l$
and $P^\gamma_y=0$ and finds explicitly
\begin{equation} \rho^\gamma_{\mu
\mu'}=\frac{1}{2}\Big((1+\mu\,P^\gamma_c)\, \delta_{\mu \mu'}-P^\gamma_l\,\delta_{\mu,
-\mu'}e^{-2i\mu\phi_\gamma}\Big)\,.
\end{equation}

Furthermore, the nucleon density matrix $\rho^p$ can be expressed in terms of irreducible
spin operators $\tau^{[I]}$ ($I=0,1$) with respect to the initial nucleon spin space,
defined in analogy to (\ref{tau-operator}),
\begin{equation}
\rho_{m_i\, {m_i}'}^p=\frac{1}{2} \sum_{I\,M}(-1)^M\,\langle \frac{1}{2}m_i|\tau^{[I]}_M
|\frac{1}{2}m_i' \rangle P^p_{I-M}\,,\label{rhoproton}
\end{equation}
where $P^p_{00}=1$, and $P^p_{1M}$ describes the sperical polarization
components of the nucleon.

We can assume that the nucleon density matrix is diagonal with respect to an orientation
axis $\vec s$ having spherical angles $(\theta_s,\phi_s)$ with respect
to the chosen coordinate system. Then one has with respect to $\vec s$
as quantization axis
\begin{equation}
\rho_{m\,m'}^p=p_m\,\delta_{m\,m'}\,,
\end{equation}
where $p_m$ denotes the probability for finding a nucleon spin projection $m$ on the
orientation axis. With respect to this axis one finds from (\ref{rhoproton})
$P^p_{I\,M}(\vec s\,)=P^p_I\,\delta_{M,0}$, where the orientation parameters $P_I^p$ are
related to the probabilities $\{p_m\}$ by
\begin{eqnarray}
P_I^p&=&\sqrt{2}\,\widehat{I}\sum_{m}(-1)^{\frac{1}{2}-m} \left(
\begin{matrix}
\frac{1}{2}&\frac{1}{2}&I \cr m &-m & 0 \cr
\end{matrix} \right)p_m\nonumber\\
&=& \delta_{I 0} + (p_\frac{1}{2}-p_{-\frac{1}{2}})\,\delta_{I 1} \,.
\end{eqnarray}
The polarization components in the chosen lab frame are obtained from the $P^p_I$ by a
rotation, transforming the quantization axis along the orientation axis into the
direction of the photon momentum, i.e.\
\beq
P^p_{IM}(\vec z\,)=P^p_Ie^{iM\phi_s}d^I_{M0}(\theta_s)\,.
\eeq
Thus the initial nucleon density matrix becomes finally
\begin{equation}
\rho_{m_i\, {m_i}'}^p=\frac{ (-1)^{\frac{1}{2}-m_i}}{\sqrt{2}} \sum_{I\,M} \left(
\begin{matrix}
\frac{1}{2}&\frac{1}{2}&I \cr m_i'&-m_i&M \cr
\end{matrix} \right) P_I^p
e^{-iM\phi_s}d^I_{M0}(\theta_s)\,. \label{rhoprotona}
\end{equation}
This means, the nucleon target is characterized by 3 parameters, namely the polarization
parameter $P_1^p$ and by the orientation angles $\theta_s$ and $\phi_s$.  If one chooses
the c.m.\ frame as reference frame, one should note that the nucleon density matrix
undergoes no change in the transformation from the lab to the c.m. system, since the
boost to the c.m.\ system is collinear with the nucleon quantization axis~\cite{Rob74}.

The evaluation of the general trace in Eq.~(\ref{trace}) can be done analogously to pion
photo production on the deuteron as described in detail in Ref.~\cite{ArF05}. In fact, one can
follow the same steps except for the use of the symmetry relation of Eq.~(12)
in~\cite{ArF05} which is different in the case of eta-pion production on the nucleon (see
Eq.~(\ref{symmetry})) because of the two pseudoscalar particles in the final state. In terms
of the small $t$-matrix elements, one finds, inserting the density
matrices of photon and nucleon, for the general trace,
\begin{eqnarray}
A_{I'M'}&=&\frac{1}{2} \sum_{\mu'\mu IM} P^p_I
\,e^{iM\phi_{\eta s}}\,d^I_{M0}(\theta_s)\,
u_{I'M';IM}^{\mu'\mu}\,\Big[(1+\mu\,P^\gamma_c)\delta_{\mu \mu'}-P^\gamma_l\,
\delta_{\mu, -\mu'}e^{2i\mu\phi_{\eta\gamma}}\Big]\,,
\end{eqnarray}
with $\phi_{\eta s}=\phi_\eta-\phi_s$ and
$\phi_{\eta\gamma}=\phi_\eta-\phi_\gamma$. Furthermore, we have
introduced the quantities
\begin{eqnarray} u_{I'M';IM}^{\mu'\mu}(q_\eta,\,
\theta_\eta,\, \theta_{\pi p},\, \phi_{pq})&=& c(q_\eta,\Omega_q,\Omega_{\pi p})\,
\widehat {I'}\widehat I\,\sum_{m_f m_f' m_i m_i'}(-1)^{m_f'-m_i} \left(
\begin{matrix}
\frac{1}{2}& \frac{1}{2}&I' \cr m_f&-m_f'&M' \cr
\end{matrix} \right)
\left(
\begin{matrix}
\frac{1}{2}& \frac{1}{2}&I \cr m_i'&-m_i&M \cr
\end{matrix} \right)\nonumber\\
&& \hspace{2cm}\times\ t^*_{m_f' \mu' m_i'}(q_\eta,\, \theta_\eta,\, \theta_{\pi p},\,
\phi_{pq}) \,t_{m_f \mu m_i}(q_\eta,\, \theta_\eta,\, \theta_{\pi p},\,
\phi_{pq})\,.\label{uim}
\end{eqnarray}
It is straightforward to prove that they behave under complex conjugation as
\begin{equation}
\label{complc} \Big(u_{I'M';IM}^{\mu'\mu}(q_\eta,\, \theta_\eta,\, \theta_{\pi p},\,
\phi_{pq})\Big)^*= (-1)^{M'+M}\,u_{I'-M';I-M}^{\mu\mu'}(q_\eta,\, \theta_\eta,\,
\theta_{\pi p},\, \phi_{pq})\,.
\end{equation}
Furthermore, with the help of the symmetry in (\ref{symmetry}) one finds
\begin{equation}
u_{I'M';IM}^{-\mu'-\mu}(q_\eta,\, \theta_\eta,\, \theta_{\pi p},\, \phi_{pq})=
(-1)^{I'+M'+I+M+\mu'+\mu}\,u_{I'-M';I-M}^{\mu'\mu}(q_\eta,\, \theta_\eta,\, \theta_{\pi
p},\, -\phi_{pq})\,,
\end{equation}
which yields in combination with (\ref{complc})
\begin{equation}\label{complca}
u_{I'M';IM}^{-\mu'-\mu}(q_\eta,\, \theta_\eta,\, \theta_{\pi p},\, \phi_{pq})=
(-1)^{I'+I+\mu'+\mu}\,\Big(u_{I'M';IM}^{\mu\mu'}(q_\eta,\, \theta_\eta,\, \theta_{\pi
p},\, -\phi_{pq})\Big)^*\,.
\end{equation}
This relation is quite useful for a further
simplification of the semi-exclusive differential cross section later on.

Separating the polarization parameters of photon ($P^\gamma_l$ and $P^\gamma_c$) and
target nucleon ($P^p_I$), it is then straightforward to show that the trace can be
brought into the form
\begin{eqnarray}
A_{I'M'}&=& \frac{1}{2}\sum_{I=0,1}P^p_I \, \sum_{M=-I}^I \,e^{iM\phi_{\eta s}}\,
d^I_{M0}(\theta_s)\,\Big[v_{I'M';IM}^1 +v_{I'M';IM}^{-1}\nonumber\\&& +
P^\gamma_c\,(v_{I'M';IM}^1-v_{I'M';IM}^{-1})
+P^\gamma_l\,(w_{I'M';IM}^1\,e^{2i\phi_{\eta\gamma}}
+w_{I'M';IM}^{-1}\,e^{-2i\phi_{\eta \gamma}})
\Big]\,,\label{trace-a}
\end{eqnarray}
where we have introduced for convenience the quantities
\begin{eqnarray}
v_{I'M';IM}^\mu(q_\eta,\, \theta_\eta,\, \theta_{\pi p},\, \phi_{pq})&=&
u_{I'M';IM}^{\mu\mu}(q_\eta,\, \theta_\eta,\, \theta_{\pi p},\, \phi_{pq})\,,\label{vim}\\
w_{I'M';IM}^\mu(q_\eta,\, \theta_\eta,\, \theta_{\pi p},\, \phi_{pq})&=& -u_{I'M';IM}^{\mu
\,-\mu}(q_\eta,\, \theta_\eta,\, \theta_{\pi p},\, \phi_{pq})\,.\label{wim}
\end{eqnarray}
According to Eqs.~(\ref{complc}) and (\ref{complca}) they have the following properties under
complex conjugation
\begin{eqnarray}
v_{I'M';IM}^\mu(q_\eta,\, \theta_\eta,\, \theta_{\pi p},\, \phi_{pq})^*&=&(-1)^{M'+M}
v_{I'-M';I-M}^\mu(q_\eta,\, \theta_\eta,\, \theta_{\pi p},\, \phi_{pq})\,,\label{v}\\
w_{I'M';IM}^\mu(q_\eta,\, \theta_\eta,\, \theta_{\pi p},\, \phi_{pq})^*&=&(-1)^{M'+M}
w_{I'-M';I-M}^{-\mu}(q_\eta,\, \theta_\eta,\, \theta_{\pi p},\, \phi_{pq})\,,\label{w}\\
v_{I'M';IM}^{\mu}(q_\eta,\, \theta_\eta,\, \theta_{\pi p},\, \phi_{pq})^*&=&(-1)^{I'+I}
v_{I'M';IM}^{-\mu}(q_\eta,\, \theta_\eta,\, \theta_{\pi p},\, -\phi_{pq}) \,,\label{vminus}\\
w_{I'M';IM}^\mu(q_\eta,\, \theta_\eta,\, \theta_{\pi p},\,\phi_{pq})^*&=&(-1)^{I'+I}
w_{I'M';IM}^\mu(q_\eta,\,\theta_\eta,\, \theta_{\pi p},\, -\phi_{pq})\,.\label{wima}
\end{eqnarray}
From Eq.~(\ref{v}) follows in particular that $v_{I'0;I0}^{\mu}$ is real.

\subsection{The differential cross section}

For the differential cross section we consider the case $I'=0$ and
$M'=0$, i.e.~$A_{00}$, for which we will use the following simplified notation
\begin{eqnarray}
v_{ IM}^\mu&=&v_{00; IM}^\mu\,,\\
w_{ IM}^\mu&=&w_{00; IM}^\mu\,.
\end{eqnarray}
The sum over $M$ in Eq.~(\ref{trace-a}) can be rearranged with the help of the relations
in Eqs.~(\ref{v}) and (\ref{w}) and $d^I_{-M0}(\theta_s)=(-1)^Md^I_{M0}(\theta_s)$
\begin{eqnarray}
\sum_{M=-I}^I \,e^{iM\phi_{\eta s}}\, d^I_{M0}(\theta_s)\,(v_{IM}^1 \pm v_{IM}^{-1})&=&
\sum_{M=0}^I \frac{d^I_{M0}(\theta_s)}{1+\delta_{M0}}\,\Big(e^{iM\phi_{\eta s}}\,
(v_{IM}^1 \pm v_{IM}^{-1})+ e^{-iM\phi_{\eta s}}\,(-1)^M
\,(v_{I-M}^1 \pm v_{I-M}^{-1})\Big)\nonumber\\
&=&\sum_{M=0}^I \frac{d^I_{M0}(\theta_s)}{1+\delta_{M0}}\, \Big(e^{iM\phi_{\gamma
s}}\,(v_{IM}^1 \pm v_{IM}^{-1})+\mbox{c.c.}\Big)\,,
\end{eqnarray}
and furthermore with
\begin{equation}\label{psiM}
\psi_M=M\phi_{\eta s}-2\,\phi_{\eta\gamma}=(M-2) \phi_{\eta}-M \phi_{s}+2\,\phi_{\gamma}\,,
\end{equation}
we get
\begin{eqnarray}
\sum_{M=-I}^I \,e^{iM\phi_{\eta s}}\, d^I_{M0}(\theta_s)\,(w_{IM}^1\,e^{-2i\phi_{\eta\gamma}} +
w_{IM}^{-1}\,e^{2i\phi_{\eta\gamma}})&=& \sum_{M=-I}^I d^I_{M0}(\theta_s)\,\Big(e^{i\psi_M}\,
w_{IM}^1 +e^{-i\psi_M}\,(-1)^M\,w_{I-M}^{-1}\Big)\nonumber\\
&=&\sum_{M=-I}^I d^I_{M0}(\theta_s)\, \Big(e^{i\psi_M}\,w_{IM}^1+\mbox{c.c.}\Big)\,.
\end{eqnarray}
This then yields for the differential cross section
\begin{eqnarray}
\frac{d\sigma}{ d\vec q_\eta d\Omega_{\pi p}}&=&
\sum_{I=0,1}P^p_I \,
\Big\{ \sum_{M=0}^I \frac{1}{1+\delta_{M0}}\,d^I_{M0}(\theta_s) \Re e \,[e^{iM\phi_{\eta
s}}\,(v_{IM}^+ +P^\gamma_c\,v_{IM}^-)] \nonumber\\&& +\,P^\gamma_l\,\sum_{M=-I}^I
d^I_{M0}(\theta_s)\, \Re e \,[e^{i\psi_M}w_{IM}^1] \Big\}\,,\label{diffcrossa}
\end{eqnarray}
where we have defined
\begin{equation}\label{vPM}
v_{IM}^\pm=v_{IM}^1\pm v_{IM}^{-1}\,.
\end{equation}
Now, introducing various beam, target and beam-target asymmetries by
\begin{eqnarray}
\tau^{0/c}_{IM}(q_\eta,\, \theta_\eta,\, \theta_{\pi p},\, \phi_{pq})&=& \frac{1}{1+\delta_{M0}}\,
\Re e \,v_{IM}^\pm(q_\eta,\, \theta_\eta,\, \theta_{\pi p},\, \phi_{pq})\,,\quad
M\ge 0\,,\label{tau0c}\\
\sigma^{0/c}_{IM}(q_\eta,\, \theta_\eta,\, \theta_{\pi p},\, \phi_{pq})&=& -\Im m \,v_{IM}^\pm(q_\eta,\,
\theta_\eta,\, \theta_{\pi p},\, \phi_{pq})\,,\quad
M> 0\,,\\
\tau^{l}_{IM}(q_\eta,\, \theta_\eta,\, \theta_{\pi p},\, \phi_{pq})&=& \Re e \,w_{IM}^1(q_\eta,\,
\theta_\eta,\, \theta_{\pi p},\, \phi_{pq})\,,\label{taul}
\\
\sigma^{l}_{IM}(q_\eta,\, \theta_\eta,\, \theta_{\pi p},\, \phi_{pq})&=& -\Im m \,w_{IM}^1(q_\eta,\,
\theta_\eta,\, \theta_{\pi p},\, \phi_{pq})\,,\label{sigmal}
\end{eqnarray}
where we took into account that $v_{I0}^{\mu}$ is real.

One obtains as final expression
for the general five-fold differential cross section including beam and target polarization
\begin{eqnarray}
\frac{d\sigma}{ d\vec q_\eta d\Omega_{\pi p}}=\frac{d\sigma_0}{ d\vec q_\eta d\Omega_{\pi p}}
\Big(1&+&P^\gamma_c\,T^{c}_{00}+P^\gamma_l\,(T^{l}_{00}\,\cos{2\phi_{\eta\gamma}}
-S^l_{00}\,\sin{2\phi_{\eta\gamma}}) \nonumber\\&&
+P^p_1 \,\Big\{\Big [T^{c}_{10}+P^\gamma_c\,T^{c}_{10}+\,P^\gamma_l\,(T^{l}_{10}\,\cos{2\phi_{\eta\gamma}}
-S^l_{10}\,\sin{2\phi_{\eta\gamma}}) \Big]\,\cos{\theta_s}\nonumber\\&&
-\frac{1}{\sqrt{2}}\,\Big[T^{0}_{10}\,\cos\phi_{\eta
  s}+S^{0}_{10}\,\sin\phi_{\eta s}
+P^\gamma_c\,\{T^{c}_{10}\,\cos\phi_{\eta
  s}+S^{c}_{10}\,\sin\phi_{\eta s}\}\nonumber\\&&
+P^\gamma_l\,\{T^{l}_{11}\,\cos(\phi_{\eta s}-2\phi_{\eta\gamma})
-T^{l}_{1-1}\,\cos(\phi_{\eta s}+2\phi_{\eta\gamma}) \nonumber\\&&
+S^{l}_{11}\,\sin(\phi_{\eta s}-2\phi_{\eta\gamma})
+S^{l}_{1-1}\,\sin(\phi_{\eta s}+2\phi_{\eta\gamma})
\}\Big]\sin\theta_s \Big\}\Big)
\,,\label{diffcross}
\end{eqnarray}
where the unpolarized differential cross section is given by
\beqa
\frac{d\sigma_0}{ d\vec q_\eta d\Omega_{\pi p}}&=&\tau^{0}_{00}\,,
\eeqa
and the various beam, target and beam-target asymmetries
\beq
T^{\alpha}_{IM}=\frac{\tau^{\alpha}_{IM}}{\tau^{0}_{00}}\,,\quad
S^{\alpha}_{IM}=\frac{\sigma^{\alpha}_{IM}}{\tau^{0}_{00}}\,,\quad\mbox{for }\alpha\in\{0,c,l\}\,.
\eeq

The corresponding derivation of the recoil polarization of the
outgoing nucleon is presented in Appendix~\ref{appa3}.

\subsection{The semi-exclusive differential cross section $\vec
p\,(\vec\gamma,\eta)\pi p$ }

We will now turn to semi-exclusive reactions where one has to integrate
over all variables which are not measured. As an example we consider the case $\vec
p\,(\vec\gamma,\eta)\pi N$ where only the produced eta is detected. This means integration
of the five-fold differential cross section $d\sigma/d\vec q_\eta d\Omega_{\pi p}$ over
$\Omega_{\pi p}$. The derivation of the resulting cross section is
presented in detail in Appendix \ref{appa2}. The cross section is
governed by the partially integrated asymmetries $\int
d\,\Omega_{\pi p}\,\tau_{IM}^\alpha$ and $\int d\,\Omega_{\pi p}\,\sigma_{IM}^\alpha$
($\alpha\in\{0,c,l\}$), of which quite a few vanish, either $\int
d\,\Omega_{\pi p}\,\tau_{IM}^\alpha$ or $\int
d\,\Omega_{\pi p}\,\sigma_{IM}^\alpha$. The final expression is
\begin{eqnarray}
\frac{d\sigma}{d\vec q_\eta}= \frac{d\sigma_0}{d\vec q_\eta}
\Big[1&+&P^\gamma_l\,\widetilde \Sigma^l\,\cos 2\phi_{\eta\gamma}
+P^p_1 \,\Big\{P^\gamma_l\,\sum_{M= -1}^1
\widetilde T_{1M}^l\sin[M\phi_{\eta s}-2\phi_{\eta\gamma}]
\,d^1_{M0}(\theta_s)\nonumber\\&& +
\sum_{M= 0}^1
\Big(-\widetilde T_{1M}^0\sin M\phi_{\eta s}
+P^\gamma_c\,\widetilde T_{1M}^c\cos M\phi_{\eta s}\Big)
\,d^1_{M0}(\theta_s)\Big\}\Big]\,,
\label{diffcrossc}
\end{eqnarray}
where the unpolarized cross section and the asymmetries are given by
\begin{eqnarray}
\frac{d\sigma_0}{d\vec q_\eta}&=&\int d\,\Omega_{\pi p}\,\tau_{00}^l(q_\eta,\, \theta_\eta,\,
\theta_{\pi p},\, \phi_{pq})=
V_{00}(q_\eta,\, \theta_\eta)\,,\label{unpoldiff}\\
\widetilde \Sigma^l(q_\eta,\, \theta_\eta)\,\frac{d\sigma_0}{d\vec q_\eta}&=&
\int d\,\Omega_{\pi p}\,\tau_{00}^l(q_\eta,\, \theta_\eta,\,
\theta_{\pi p},\, \phi_{pq})= W_{00}(q_\eta,\, \theta_\eta)\,,\label{sigasy}\\
\widetilde T_{1M}^0(q_\eta,\, \theta_\eta)\,\frac{d\sigma_0}{d\vec q_\eta}&=&
\int d\,\Omega_{\pi p}\,\sigma_{1M}^0(q_\eta,\, \theta_\eta,\,
\theta_{\pi p},\, \phi_{pq})=-(2-\delta_{M0})\, \Im m\, [V_{1M}(q_\eta,\,
\theta_\eta)]\,,\quad\mbox{for }M=0, 1\,,
\label{tim}\\
\widetilde T_{1M}^c(q_\eta,\, \theta_\eta)\,\frac{d\sigma_0}{d\vec q_\eta}&=&
\int d\,\Omega_{\pi p}\,\tau^c_{1M}(q_\eta,\, \theta_\eta,\,
\theta_{\pi p},\, \phi_{pq})= (2-\delta_{M0})\, \Re e\, [V_{1M}(q_\eta,\,
\theta_\eta)]\,,\quad\mbox{for }M=0, 1\,,
\label{timc}\\
\widetilde T_{1M}^l(q_\eta,\, \theta_\eta)\,\frac{d\sigma_0}{d\vec q_\eta}&=&
\int d\,\Omega_{\pi p}\,\sigma_{01M}^l(q_\eta,\, \theta_\eta,\,
\theta_{\pi p},\, \phi_{pq})= i\,W_{1M}(q_\eta,\, \theta_\eta)\,, \quad\mbox{for }M=0,\pm 1\,.\label{timl}
\end{eqnarray}
Here, the quantities $V_{IM}$ and $W_{IM}$ are
related to the small $v^1_{IM}$ and $w^1_{IM}$ by
\begin{eqnarray}
V_{IM}(q_\eta,\, \theta_\eta)&=&\int d\,\Omega_{\pi p}\,v_{IM}^1(q_\eta,\, \theta_\eta,\,
\theta_{\pi p},\, \phi_{pq})\,,\\
 W_{IM}(q_\eta,\, \theta_\eta)&=& \int d\,\Omega_{\pi p}\,w_{IM}^{1}(q_\eta,\,
\theta_\eta,\, \theta_{\pi p},\, \phi_{pq})\,.
\end{eqnarray}
Because $V_{I0}$ is real according to Eq.~(\ref{v}), the asymmetries $\widetilde T_{00}^c$
and $\widetilde T_{10}^0$ vanish identically. Furthermore, one should
note that $W_{1M}$ is purely imaginary. This is shown in Appendix~A
(see Eq.~(A7)).  More explicitly one has
\beqa
\frac{d\sigma}{d\vec q_\eta}= \frac{d\sigma_0}{d\vec q_\eta}
\Big[1&+&P^\gamma_l\,\widetilde \Sigma^l\,\cos 2\phi_{\eta\gamma}
+P^p_1 \,\Big\{P^\gamma_l\,\Big(-\widetilde T_{10}^l\cos \theta_s\cos
2\phi_{\eta\gamma}\nonumber\\&&
-\frac{1}{\sqrt{2}}\,[(\widetilde T_{1-1}^l+\widetilde T_{11}^l)
\sin \phi_{\eta s}\cos 2\phi_{\eta\gamma}
+(\widetilde T_{1-1}^l-\widetilde T_{11}^l)
\cos \phi_{\eta s}\sin 2\phi_{\eta\gamma}]\sin \theta_s\Big)
\nonumber\\&&
+\frac{1}{\sqrt{2}}\,\widetilde T_{11}^0\sin \phi_{\eta
  s}\sin \theta_s
-P^\gamma_c\,\Big(\widetilde T_{11}^c\cos \phi_{\eta s}\sin \theta_s
-\frac{1}{\sqrt{2}}\,\widetilde T_{10}^c\cos \theta_s\Big)\Big\}\Big]
\,,
\label{diffcrossd}
\eeqa

We would like to point out that in forward
and backward eta emission, i.e.\ for $\theta_\eta=0$ and $\pi$, the following asymmetries
have to vanish
\begin{equation} \widetilde \Sigma^l=0\,,\quad \widetilde
T_{11}^{0,c}=0\,\quad\mbox{and}\quad  T_{1M}^{l}=0\,,
\label{asym0}
\end{equation}
because in that case the differential cross section cannot depend on $\phi_\eta$, since
at $\theta_\eta=0$ or $\pi$ the azimuthal angle $\phi_\eta$ is undefined or arbitrary.
This feature can also be shown by straightforward evaluation of $V_{IM}$ and $W_{IM}$
using the explicit representation of the $T$-matrix in Eq.~(\ref{smallt}). One finds
\begin{equation}
V_{IM}(q_\eta,\, \theta_\eta=0/\pi,\, \theta_{\pi p},\, \phi_{pq})=0 \quad \mbox{for
}\,M\neq 0\,,\quad W_{IM}(q_\eta,\, \theta_\eta=0/\pi,\, \theta_{\pi p},\,
\phi_{pq})=0 \quad\mbox{for all }M\,.\label{theta0}
\end{equation}
The formulas above can readily be extended to the other cases of an
active pion or proton through a simple replacement of the appropriate angles with a
corresponding redefinition of the various planes in Fig.~1.

\subsection{The total cross section}

The general total cross section is obtained from Eq.~(\ref{diffcrossc}) by integrating over
$d^3q_\eta$ resulting in
\begin{equation}
\sigma = \sigma_0\,\Big[1 +P^\gamma_c\,P^p_1\,\overline T_{10}^{\,c}\,\cos\theta_s
\Big]\,,
\end{equation}
where the unpolarized total cross section and the only beam-target
asymmetry $\overline T_{10}^{\,c}$ are given by
\begin{eqnarray}
\sigma_0&=&2\pi\int d\cos\theta_\eta\int_{q_\eta^{min}}^{q_\eta^{max}}
q^2_\eta dq_\eta\,\frac{d\sigma_0}{d\vec q_\eta}\,,\\
\sigma_0\,\overline T_{10}^{\,c}&=&2\pi\int d\cos\theta_\eta\int_{q_{min}}^{q_{max}}
q^2dq_\eta\,\frac{d\sigma_0}{d\vec q_\eta}\,T_{10}^{\,c}\,.
\end{eqnarray}
The integration limits $q_\eta^{min}$ and $q_\eta^{max}$ are
determined by energy and momentum conservation.
There is no dependence on the linear photon polarization as expected.

\section{Results and discussion}
\label{results}

In this section we present our results for those asymmetries of the reaction $\gamma
p\to\pi^0\eta p$ for which data already exist or are expected to be measured in the near
future. The observables are calculated in the overall $\gamma p$ c.m.\ frame. The main
ingredients of our model are described in detail in Ref.~\cite{FOT,FKLO}. Here we limit
ourselves to a brief overview of the model needed for the discussion. The calculation is
based on a conventional isobar model as used, for example, for double pion
photoproduction in Refs.~\cite{Oset,Laget,Ochi,Mokeev,FA}. The model parameters were
fitted to the angular distributions of pions in the $\pi p$ c.m.\ system, as well as to
the distribution over the polar angle of $\eta$ in the overall c.m.\ frame. The
corresponding data were presented in Refs.~\cite{Kashev} and \cite{FKLO} in the region up
to a photon lab energy $\omega_\gamma=1.4$~GeV. The present results are obtained in the
same energy interval.

The reaction amplitude comprises background and resonance terms
\begin{equation}\label{15}
t_{m_f\lambda}=t^B_{m_f\lambda}+\sum_{R(J^\pi;T)}t^R_{m_f\lambda}\,.
\end{equation}
An individual resonance state $R(J^\pi;T)$ is determined by its spin-parity $J^\pi$ and
isospin $T$. Instead of the spin projections  of the initial particles
$m_i$ and $\mu$, respectively,  we use their
sum $\lambda=m_i+\mu=\pm 1/2$, $\pm 3/2$, which in our coordinate system with the
quantization axes along the incident photon momentum corresponds to the initial state helicity.

The resonance sector includes only states with isospin $T=3/2$. As already noted,
analysis of the existing data for $\gamma p\to\pi^0\eta p$ are in agreement with the
assumption that in the energy region studied here the reaction is
dominated by the $D_{33}$ wave. In
the present model the latter is populated by the $D_{33}(1700)$ and $D_{33}(1940)$
states. The one-star resonance $D_{33}(1940)$ was first introduced
into the reaction $\gamma p\to\pi^0\eta p$ in Ref.~\cite{Horn} based
on a partial wave analysis (PWA). In our model the status of this
baryon is still not very clear. Primarily we need it in order to maintain
the importance of the $D_{33}$ wave at energies above 1.3~GeV, which otherwise would
rapidly decrease with increasing energy. Other $T=3/2$ resonances entering the amplitude are
$P_{33}(1600)$, $P_{31}(1750)$, $F_{35}(1905)$, and $P_{33}(1920)$. Their parameters
resulting from a fit are listed in Table~II of Ref.~\cite{FKLO}.

As is shown in Refs.~\cite{Dor,FKLO}, the background contribution is
small, so that we can focus our attention on the resonance sector
alone. According to the isobar model concept each
resonance term is given by a coherent sum of individual amplitudes corresponding to
intermediate transitions to $\eta\Delta(1232)$ and $\pi S_{11}(1535)$ configurations
\begin{equation}\label{15a}
t^R_{m_f\lambda}=t^{R(\eta\Delta)}_{m_f\lambda}+t^{R(\pi N^*)}_{m_f\lambda}\,,
\end{equation}
where the resonances $\Delta(1232)$ and $S_{11}(1535)$ are denoted as $\Delta$ and $N^*$,
respectively.
The $\pi\eta$ system is assumed not to resonate in our energy
interval. The validity of this
assumption is confirmed by the results of Ref.~\cite{Horn} where the contribution of the
resonance $a_0(980)$ at energies $\omega_\gamma<1.4$~GeV is shown to be less than
1~$\%$.

Each term in Eq.~(\ref{15a}) has the form
\begin{eqnarray}\label{20}
&&t_{m_f\lambda}^{R(\alpha)}(W,\vec{q}_\pi,\vec{q}_\eta,\vec{p}_f)=
c_R^{(\alpha)}A^R_\lambda\, f^{R(\alpha)}_{m_f\lambda}
(\vec{q}_\pi,\vec{q}_\eta,\vec{p}_f),\quad  \alpha\in\{\eta\Delta,\,\pi N^*\}\,,
\end{eqnarray}
with $W$ being the total c.m.\ energy. The quantities $A_\lambda^R$, which in
general depend on $W$, are helicity functions determining the transition $\gamma
p\to R$. The factor $c_R^{(\alpha)}$ absorbs all quantities which are independent of the
quantum numbers $m_f$ and $\lambda$. Its exact form is irrelevant for the formalism to
follow. The angular dependent part $f^{R(\alpha)}_{m_f\lambda}$ describes the decay of
the resonance $R$ into $\pi\eta N$ via intermediate formation of an $\eta\Delta$ or $\pi
N^*$ state.

In the actual calculation we adhere to the non-relativistic concept of angular momentum
so that the angular dependence of the amplitudes (\ref{20}) is described by means of
spherical harmonics
\begin{eqnarray}\label{35a}
&&f^{R(\eta\Delta)}_{m_f\lambda}\sim \sum_{m,m_\eta,m_\Delta} \left(\begin{array}{ccc} 1 & \frac12 &
\frac32\cr m & m_f & -m_\Delta\cr
\end{array}\right)
\left(\begin{array}{ccc} l_\eta & \frac32 & J\cr m_\eta & m_\Delta & -\lambda\cr
\end{array}\right)
\,Y_{1m}(\Omega_{\pi p})\,d^{l_\eta}_{m_\eta 0}(\theta_\eta)\,,\\
\label{35b} &&f^{R(\pi N^*)}_{m_f\lambda}\sim \sum_{m_\pi}
\left(\begin{array}{ccc} l_\pi & \frac12 &
J\cr m_\pi & m_f & -\lambda\cr
\end{array}\right)\,
Y_{l_\pi m_\pi}(\Omega_{\pi})\nonumber\\
&&\phantom{xxxxxx}\sim \sum_{m_\pi}
\left(\begin{array}{ccc} l_\pi & \frac12 & J\cr m_\pi & m_f &
-\lambda\cr \end{array}\right)\, \sum\limits_{l=0}^{l_\pi}A_l^{l_\pi} \sum_m
\left(\begin{array}{ccc} l_\pi-l & l & l_\pi \cr m_\pi-m & m & -m_\pi\cr
\end{array}\right)\,
Y_{lm}(\Omega_{\pi p})\,d^{l_\pi-l}_{m_\pi-m\,0}(\theta_\eta)\,.
\end{eqnarray}
The coefficients $A_l^{l_\pi}$, determined as
\begin{equation}
A_l^{l_\pi}=\left(\frac{m_\pi
q_\eta}{(m_\pi+M_p)p_{\pi p}}\right)^l\sqrt{\frac{(2l_\pi-1)\,(2l_\pi)!}{(2l-1)\,(2l_\pi-2l)!\,(2l)!}}\,,
\end{equation}
stem from the expansion of the function $Y_{l_\pi m_\pi}(\Omega_\pi)$
with respect to products of spherical functions depending on
$\Omega_{\pi p}$ and $\Omega_\eta$.

\begin{figure}
\includegraphics[scale=.8]{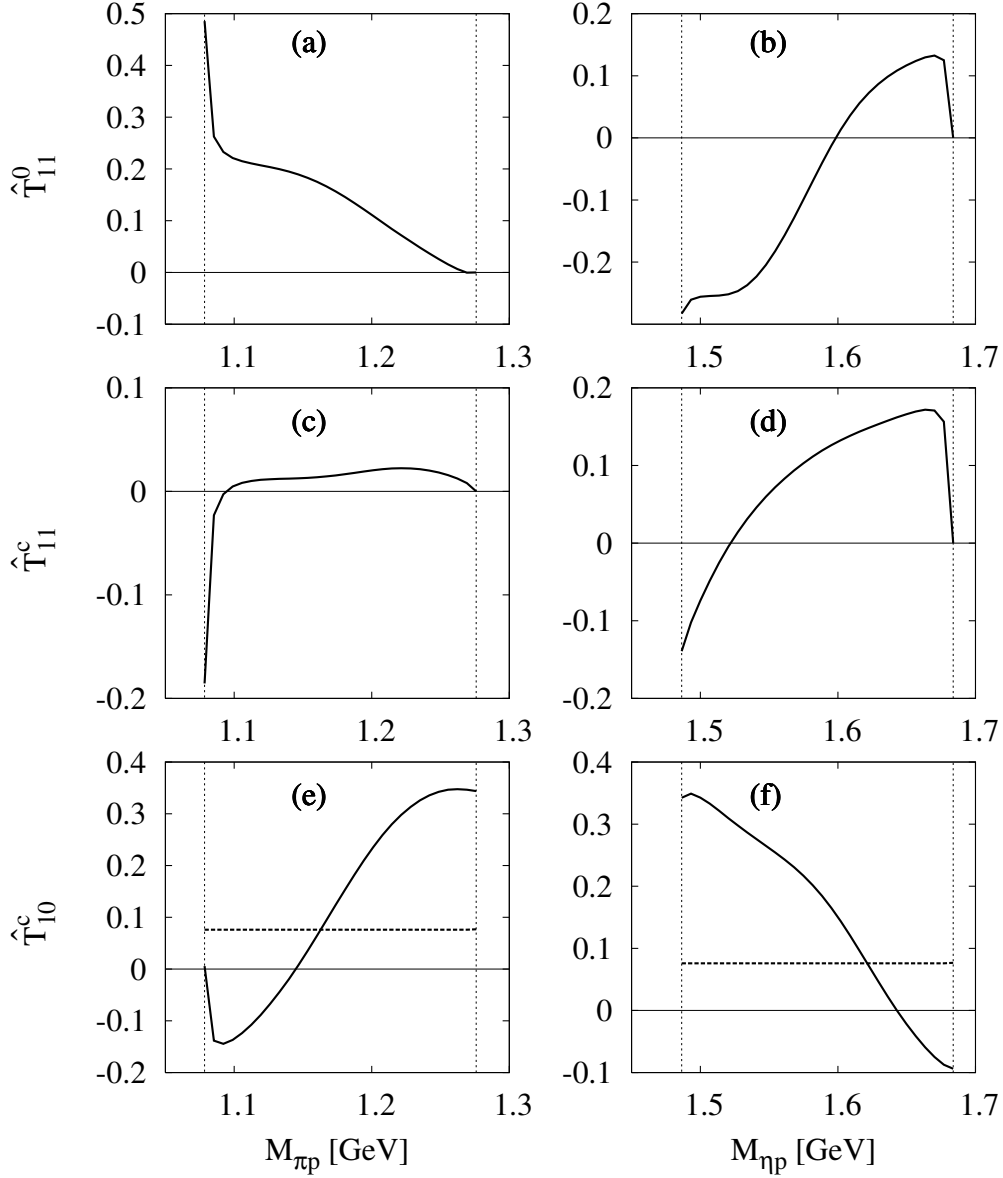}
\caption{The target asymmetry $\widehat  T^0_{11}$ ((a) and (b)), and the beam-target
  asymmetries for circularly polarized photons
  $\widehat  T^c_{11}$ ((c) and (d)) and $\widehat  T^c_{10}$ ((e) and
  (f))  for the $\eta$ as active particle (left panels)  as function of the
  $\pi p$ invariant mass spectrum $M_{\pi p}$, and for the pion as
  active particle (right panels) as function of $M_{\eta p}$, calculated at a lab photon
  energy of 1.3~GeV. The solid line presents the full calculation. The
  dashed line is obtained including the $D_{33}(1700)$ resonance
  only. The asymmetries $\widehat  T^0_{11}$ and $\widehat  T^c_{11}$ vanish in the single
  $D_{33}$ model. The vertical dotted lines mark the boundaries of the
  available kinematical region.} \label{figTFE}
\end{figure}

\begin{figure}
\begin{center}
\includegraphics[scale=.8]{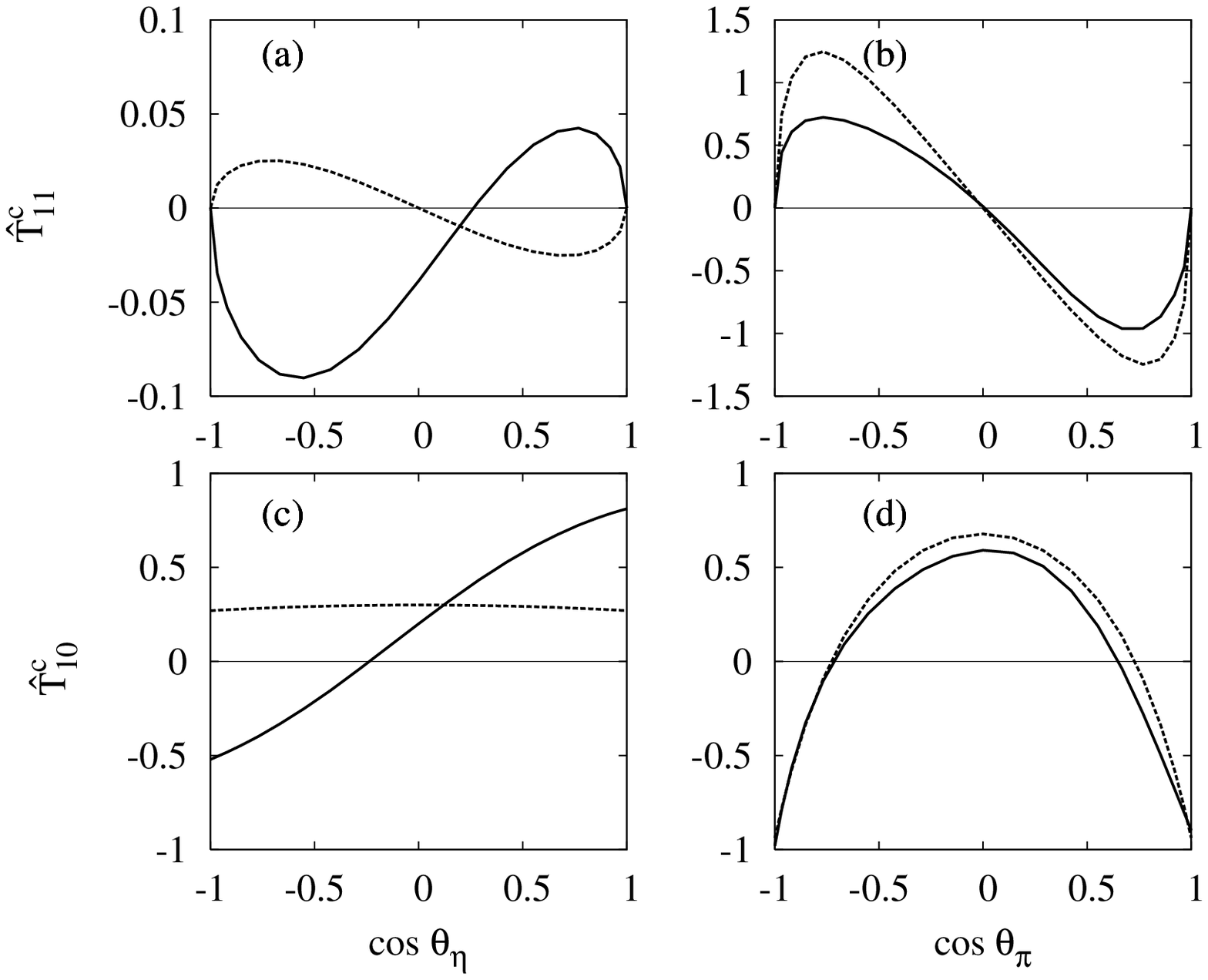}
\caption{The beam-target asymmetries for circularly polarized photons $\widehat
  T^c_{11}$ and $\widehat  T^c_{10}$ at a lab photon
  energy of 1.3 GeV as function of the polar angles of active eta
  (left panels (a) and (c)) and active pion (right panels (b) and (d)) in the
  $\gamma p$ c.m.\ frame. The solid curve is the full model
  calculation. The dashed curve includes only the $D_{33}(1700)$ and
  $D_{33}(1940)$ resonances.}
\label{figFE}
\end{center}
\end{figure}

\subsection{The semi-exclusive asymmetries for circularly polarized photons and
  polarized protons}

Now we will turn to the case where the active particle ($\pi,\eta$
or $p$) is 
measured for a fixed invariant mass of the other two final particles
irrespective of the direction $\theta_{\alpha}$ with $\alpha=\pi,\eta$
or $p$, respectively, for a fixed reaction plane. The resulting
semi-exclusive differential cross section is obtained by  an
additional integration over the polar angles $\theta_\alpha$,
respectively. It is given by an expression formally analogous to
Eq.~(\ref{diffcrossc}) with the following replacements (for the eta as
active particle as example)
\begin{eqnarray}
\frac{d\sigma_0}{d\vec q_\eta}\,&\rightarrow&\frac{d\sigma_0}{dM_{\pi p}d\phi_{\eta}}=\int
d\cos\theta_\eta\, {\cal K}\,\frac{d\sigma_0}{d\vec q_\eta}\,,\label{hat-asymmetries_a}\\
\frac{d\sigma_0}{d\vec q_\eta}\,\widetilde \Sigma^l(q_\eta,\, \theta_\eta)
&\rightarrow&\frac{d\sigma_0}{dM_{\pi p}d\phi_{\eta}}\,
\widehat \Sigma^l(M_{\pi p})=\int d\cos\theta_\eta\,
{\cal K}\,\frac{d\sigma_0}{d\vec q_\eta}\,\widetilde \Sigma^l(q_\eta,\, \theta_\eta)\,,\label{hat-asymmetries_b}\\
\label{TimLq} \frac{d\sigma_0}{d\vec q_\eta}\,\widetilde T_{IM}^\alpha(q_\eta,\,
\theta_\eta) &\rightarrow&\frac{d\sigma_0}{dM_{\pi p}d\phi_{\eta}}\,
\widehat  T^\alpha_{IM}(M_{\pi p})=\int
d\cos\theta_\eta\,{\cal K}\,\frac{d\sigma_0}{d\vec q_\eta}\, \widetilde
T^\alpha_{IM}(q_\eta,\, \theta_\eta)\,,\quad \alpha\in\{0,l,c\}\,.\label{hat-asymmetries_c}
\end{eqnarray}
The factor ${\cal K}$ takes into account the transformation of the
differential, i.e.\ $q_\eta^2dq_\eta={\cal K}\,dM_{\pi p}$. In the $\gamma p$
c.m.\ frame, this factor is independent of $\theta_\eta$ and reads
\begin{equation}
{\cal K}=\frac{q_\eta\omega_\eta M_{\pi p}}{W}\,.
\end{equation}
For the case of an active pion or proton one simply has to make the
following replacements: $\eta\to\pi$~or~$p$ and $\pi p \to \eta
p$~or~$\pi\eta$, respectively.

We now consider circularly polarized photons and allow for polarized
protons, i.e.\ $P^\gamma_c\neq 0$, $P^p_1\neq 0$, and $P^\gamma_l= 0$.
Furthermore, we set the azimuthal $\eta$ angle to $\phi_\eta=0$. Then one obtains explicitly
\beqa
\frac{d\sigma}{dM_{\pi
    p}d\phi_{\eta}}\Big|_{\phi_\eta=0}&=&\frac{d\sigma_0}{dM_{\pi
    p}d\phi_{\eta}}\Big|_{\phi_\eta=0} 
\Big[1+P_1^p\Big(\frac{1}{\sqrt{2}}\, \widehat  T^0_{11}\sin{\phi_{s}}
\sin{\theta_s} +P_c^\gamma ( \widehat  T^c_{10}\cos{\theta_s}
-\frac{1}{\sqrt{2}}\, \widehat  T^c_{11}\cos{\phi_{s}}\sin{\theta_s})\Big)\Big]
\eeqa
As a sideremark, angular distributions irrespective of the energy of
the active particle may be obtained in a similar manner via
appropriate  integration of the cross section in~(\ref{diffcrossc})
over the energy of the active particle.

Figure~\ref{figTFE} demonstrates our predictions for the
semi-inclusive target asymmetry $\widehat  T^0_{11}$ as well as
for the double polarization observables $\widehat  T^c_{11}$ and $\widehat  T^c_{10}$.
In the single $D_{33}$ resonance model including only $D_{33}(1700)$
and $D_{33}(1940)$, both  asymmetries $\widehat  T^0_{11}$ and $\widehat  T^c_{11}$ should
vanish completely. The corresponding angular distributions (in
Fig.~\ref{figFE} we show the dependence of $\widehat  T^c_{11}$ on
$\cos\theta_\eta$ and $\cos\theta_\pi$)
are odd functions of $\cos\theta_{\eta/\pi}$, so that they vanish after integration over the polar
angle. The full model, in which also positive parity resonances are included, gives an
even component in both asymmetries thus leading to a rather intricate energy dependence as
is shown in Fig.~\ref{figTFE}.

It is also worth noting, that for the active pion the dependence of $\widehat  T^c_{11}$ on
$\theta_\pi$ is rather similar to that
observed for single $\pi^0$ photoproduction in the $\Delta$ region. This may be due to
the dominance of the $s$ wave in the $\eta\Delta$ channel and to the
relatively large $\eta$ mass, so that the $\Delta$ decay is not
contaminated by the presence of an $\eta$ meson.

Of special interest is the observable $\widehat  T^c_{10}$. In the
single $D_{33}$ model its value is
almost independent of $M_{\pi p}$ (or $M_{\eta p}$). For example, if
only the $D_{33}(1700)$ resonance is retained in the amplitude it is
approximately equal to
\begin{equation}\label{35}
\widehat  T^c_{10}\approx \frac{1-a^2}{2(1+a^2)}\,,\quad\mbox{with } a=\frac{A_{3/2}}{A_{1/2}}\,,
\end{equation}
where $A_{\lambda}$ is a helicity function corresponding to the transition $\gamma N\to
D_{33}(1700)$ (see our ansatz (\ref{20}) for the resonance amplitudes). Taking $a=1.1$ from
the analysis of Ref.~\cite{Kashev} (see Fig.~6 of~\cite{Kashev} at $E_\gamma=1.3$ GeV),
we will have, according to Eq.~(\ref{35}), $\widehat  T^c_{10}=-0.05$ in general agreement with the
result shown by the dashed line in Fig.~\ref{figTFE}. If both resonances $D_{33}(1700)$
and $D_{33}(1940)$ are included, $\widehat  T^c_{10}$ remains constant, but its value is no longer
determined by a simple relation analogous to (\ref{35}). As we can see, inclusion of
other resonances, resulting in a strong interference with the leading partial wave, crucially
changes the shape of $\widehat  T^c_{10}$.

It is also interesting to note that in contrast to single pseudoscalar meson photoproduction
$\widehat  T^c_{10}$ does not approach unity at very forward and
backward $\eta$ angles (see panel (c) in Fig.~\ref{figFE}). The reason
for this behaviour lies in the spin $3/2$ of the $\Delta$ 
resonance, so that angular momentum conservation does not require $\lambda=1/2$ at
$\theta_\eta=0(\pi)$, as in the case of a single meson.

\begin{figure}
\begin{center}
\includegraphics[scale=.8]{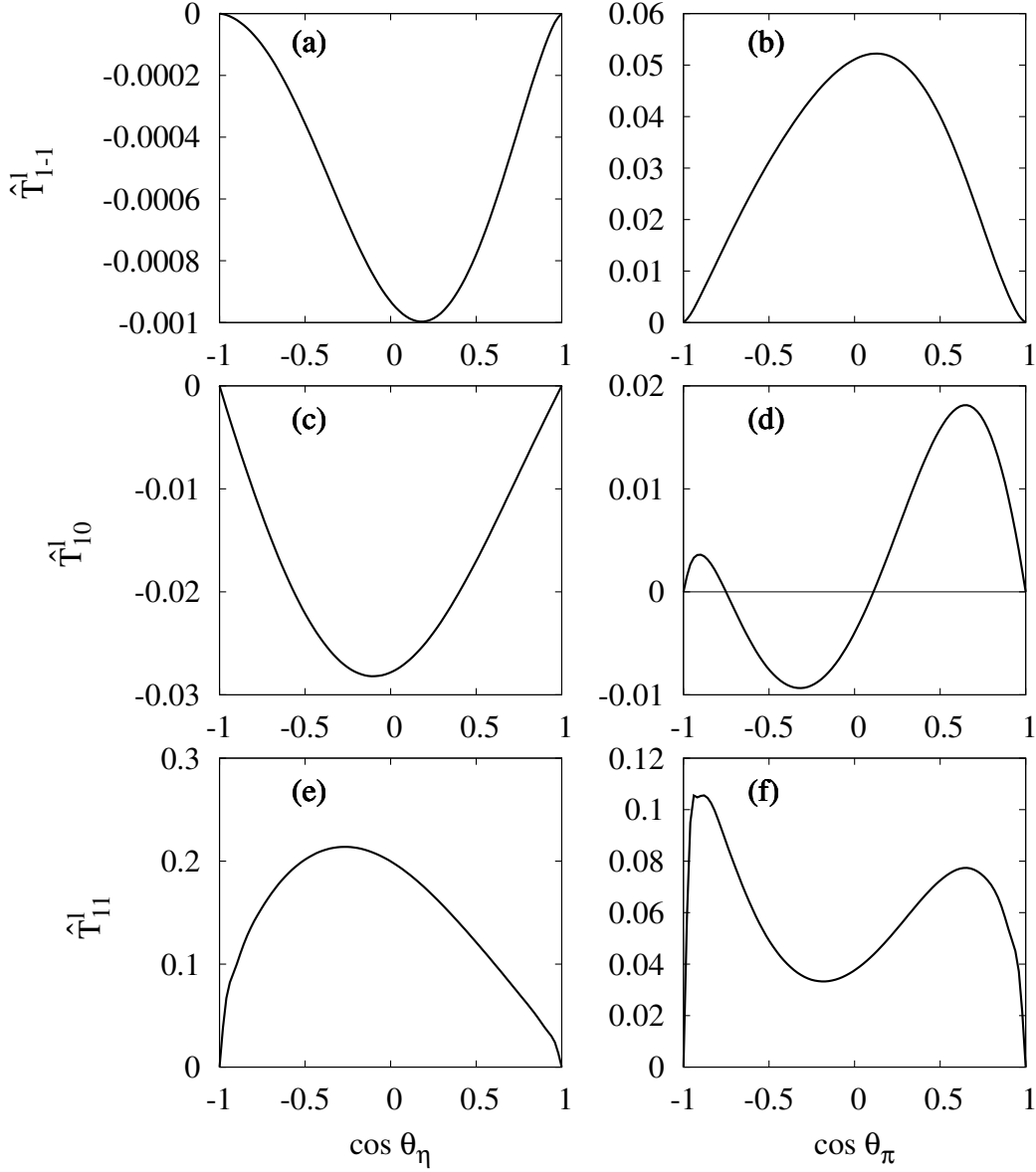}
\caption{The beam-target asymmetries for linearly polarized photons $\widehat
  T^l_{1-1}$, $\widehat  T^l_{10}$,
  and $\widehat  T^l_{11}$ at a lab photon
  energy of 1.3 GeV as function of the  polar angles of an active eta
  (left panels (a), (c), and (e)) and an active pion (right panels
  (b), (d), and (f)) in the
  $\gamma p$ c.m.\ frame.} \label{linear-beam-target-asy}
\end{center}
\end{figure}

\subsection{The semi-exclusive asymmetries for linearly polarized photons and
  polarized protons}

For only linearly polarized photons the semi-exclusive cross section is again obtained
from Eq.~(\ref{diffcrossc}) for $P^\gamma_c=0$ and $\phi_\eta=0$ with the replacements of
Eqs.~(\ref{hat-asymmetries_a}) through  (\ref{hat-asymmetries_c})
\begin{eqnarray}
\frac{d\sigma}{dM_{\pi p}d\phi_{\eta}}\Big|_{\phi_\eta=0}&=&\frac{d\sigma_0}{dM_{\pi
p}d\phi_{\eta}}\Big|_{\phi_\eta=0} \Big[1+P^\gamma_l\,\Big\{\widehat \Sigma^l\,\cos
2\phi_{\gamma} +P^p_1 \,\Big(-\widehat T_{10}^l\cos \theta_s\cos
2\phi_{\gamma}\nonumber\\ && +\frac{1}{\sqrt{2}}\,\Big[(\widehat T_{1-1}^l+\widehat
T_{11}^l) \sin \phi_{s}\cos 2\phi_{\gamma} +(\widehat T_{1-1}^l-\widehat T_{11}^l) \cos
\phi_{s}\sin 2\phi_{\gamma}\Big]\sin \theta_s\Big) \Big\}\Big] \,,
\end{eqnarray} where
$\phi_{\gamma}$ measures the angle between the reaction and the photon plane. The gross
features of the beam asymmetry for linearly polarized photons $\widehat \Sigma^l$ as a
function of the $\pi N$ or $\eta N$ invariant energies were already discussed in detail
in Ref.~\cite{FKLO}. Therefore, we show here only the additional beam-target
asymmetries $\widehat  T^l_{1M}$ in Fig.~\ref{linear-beam-target-asy}.

\begin{figure}[htb]
\includegraphics[scale=.8]{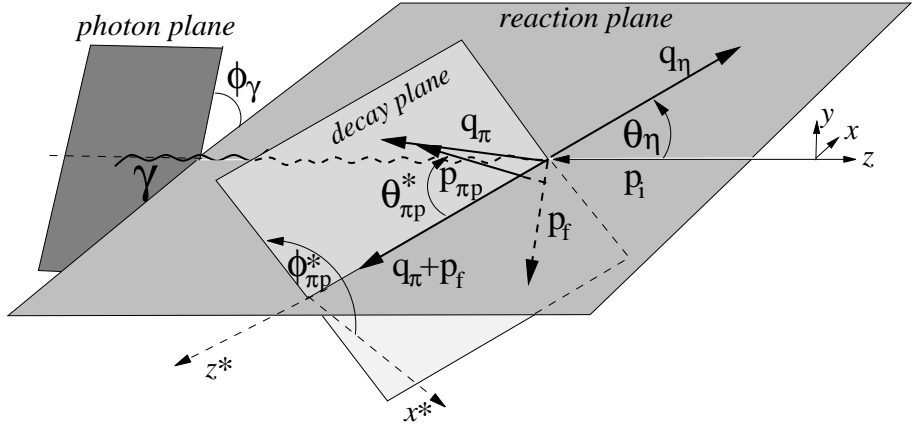}
\caption{Kinematics of $\pi\eta$ photoproduction on the nucleon for an
  active eta in the
  c.m.\ system with rotatet coordinate system. }
\label{fig_kinematics_cm}
\end{figure}

Furthermore, we present results for the asymmetries called $I^c$ and
$I^s$ which were recently
measured at ELSA \cite{Gutz_Is}. In this experiment the direction of
the eta meson was detected in the reaction plane in coincidence with
the pion proton pair for a fixed orientation of the decay plane
integrated over the direction within this plane of $\vec p_{\pi p}$ as
function of the angle between the
reaction plane and the decay plane.  The initial proton was unpolarized.
For the comparison of our results with the data we have adjusted
the calculation to the experimental kinematic conditions of these
measurements. First of all, we
changed the coordinate system as defined in
Fig.~\ref{fig_kinematics} for the $x$-$z$-plane coinciding with the
reaction plane, i.e.\ $\phi_\eta=0$
($z$-axis parallel to $\vec k$ and $y$-axis parallel to $\vec k
\times\vec q_\eta$) by rotating it around the $y$-axis such that the
new $z^*$-axis is aligned along the vector $\vec{q}_\pi+\vec{p}_f$.
With respect to the rotated coordinate system the relative momentum
$\vec p_{\pi p}$ has the spherical angles $\Omega_{\pi p}^*=(\theta_{\pi p}^*,\phi_{\pi p}^*)$, and the decay plane intersects the
reaction plane with the azimuthal  angle $\phi_{\pi p}^*$. This is illustrated in
Fig.~\ref{fig_kinematics_cm} for the c.m.\ system.
 In the rotated $\gamma p$
c.m.\ system the corresponding expressions for the amplitudes
$f^{R(\alpha)}_{m_f\lambda}$ can easily be obtained from eqs.~(\ref{35a}) and
(\ref{35b}) via a positive rotation of $Y_{lm}(\Omega_p)$ by an angle
$\theta_R=\theta_\eta+\pi$ around the $y$-axis.  With respect to the new
variables one obtains a set of new structure functions
$\tau/\sigma^{(*)\alpha} _{IM}(q_\eta,\, \theta_\eta,\, \theta_{\pi p}^*,\,
  \phi_{pq}^*)$ which are related to the old ones by the Jacobian
\beq
J(\cos\theta_{\pi p},\phi_{\pi p};\cos\theta_{\pi p}^*,\phi_{\pi
p}^*)=\Big|\frac{\partial(\cos\theta_{\pi p},\phi_{\pi p})}{\partial(\cos\theta_{\pi p}^*,\phi_{\pi
p}^*)}\Big|
\eeq
according to
\begin{equation}
\tau/\sigma^{(*)\alpha}_{IM} (q_\eta,\, \theta_\eta,\, \theta^*_{\pi p},\,
  \phi^*_{pq})=\tau/\sigma^{\alpha}_{IM} (q_\eta,\, \theta_\eta,\, \theta_{\pi p},\,
  \phi_{pq})\,|J(\cos\theta_{\pi p},\phi_{\pi p};\cos\theta_{\pi p}^*,\phi_{\pi
p}^*)|\quad\mbox{for }\alpha\in\{0,l,c\} \,.\label{obs-star}
\end{equation}
From the relations between $(\theta_{\pi p},\phi_{\pi p})$ and
$(\theta^*_{\pi p},\phi^*_{\pi p})$ 
\begin{eqnarray}
\cos\theta_{\pi p}&=&\cos\theta^*_{\pi p}\cos\theta_R-\sin\theta^*_{\pi p}\cos\phi_{\pi
p}\sin\theta_R\,,\\
\cot\phi_{\pi p}&=&\cot\phi^*_{\pi p}\cos\theta_R
+\frac{\cot\theta^*_{\pi p}}{\sin\phi^*_{\pi p}}\,\sin\theta_R
\end{eqnarray}
with $\theta_R$ denoting the rotation angle, one can see that $\theta_{\pi p}$ and
$\phi_{\pi p}$ are, respectively, even and odd functions of $\phi^*_{\pi p}$ (what may
also be obvious from the geometric considerations). Explicitely, one finds for the Jacobian
\begin{equation}\label{Jac}
J(\cos\theta_{\pi p},\phi_{\pi p};\cos\theta_{\pi p}^{\ast},\phi_{\pi p}^*)=
\frac{\sin^2\phi_{\pi p}}{\sin^2\theta^{\ast}_{\pi p}\sin^2\phi_{\pi p}^*}
\Big((\sin\theta_{\pi p}^*\cos\theta_R
+\cos\phi^*_{\pi p}\cos\theta_{\pi p}^*\sin\theta_R)^2+\sin^2\phi_{\pi p}^*\sin^2\theta_R\Big)\,.
\end{equation}
The above found symmetry of the angle transformation is reflected by 
the invariance of the Jacobian under
simultaneous sign change of $\phi_{\pi p}$ and $\phi_{\pi p}^*$, i.e. 
\begin{equation}\label{Jacsym}
J(\cos\theta_{\pi p},-\phi_{\pi p};\cos\theta_{\pi p}^*,-\phi_{\pi
p}^*) =J(\cos\theta_{\pi p},\phi_{\pi p};\cos\theta_{\pi p}^*,\phi_{\pi
p}^*)\,,
\end{equation}
which will be used later on.

As mentioned in the formal part, we also consider partitions
$\pi+(\eta N)$ and $p+(\pi\eta)$ in which the decay plane is spanned
in the former case by the vectors $\vec q_\eta$ and $\vec p_f$, formally
replacing $\Omega^*_{\pi p}$ by $\Omega^*_{\eta p}=(\theta^*_{\eta
  p},\phi^*_{\eta p})$, and in the latter case by $\vec q_\pi$ and
$\vec q_\eta$ with $\Omega^*_{\pi\eta}=(\theta^*_{\pi\eta},\phi^*_{\pi\eta
})$.

In order to evaluate the corresponding semi-exclusive
observables one has to integrate over $dq_\eta$ and $d\cos\theta_{\pi
  p}^*$ the general expression for the
differential cross section, which reads for $\phi_\eta=0$,
$P^\gamma_c=0$, and $P^p_1=0$
\beq
\frac{d\sigma}{ d\vec q_\eta d\Omega^*_{\pi p}}=
\frac{d\sigma_0}{ d\vec q_\eta d\Omega^*_{\pi p}}
\Big(1+P^\gamma_l\,(T^{(*)l}_{00}\,\cos{2\phi_{\gamma}}
+S^{(*)l}_{00}\,\sin{2\phi_{\gamma}}) \Big)\,.\label{diffcross-star}
\eeq
This then yields in the
notation of  Ref.~\cite{Roberts}  (one
should note that $\phi'$ in Ref.~\cite{Gutz_Is} is related to
$\phi_{\gamma}$ by $\phi_{\gamma}=\phi'-\pi/2$)
\beq
\frac{d\sigma}{ d\phi_\eta d\phi_{\pi p}^*}
=\frac{d\sigma_0}{ d\phi_\eta d\phi_{\pi p}^*}
\Big(1-P^\gamma_l\,(I^c(\phi_{\pi p}^*)\,\cos{2\phi_{\gamma}}
+I^s(\phi_{\pi p}^*)\,\sin{2\phi_{\gamma}}) \Big)\,,
\eeq
where the linear beam asymmetries $I^c$ and $I^s$ are determined by the coefficients
$S^{(*)l}_{00}$ and $T^{(*)l}_{00}$ in Eq.~(\ref{diffcross-star})
\begin{eqnarray}
I^c(\phi^*_{\pi p})\frac{d\sigma}{d\phi_\eta d\phi^*_{\pi p}}&=& -\int d\cos\theta_{\pi
p}^*\int_{q_\eta^{min}}^{q_\eta^{max}} q^2_\eta dq_\eta\,\frac{d\phi_\eta
d\sigma_0}{d\phi^*_{\pi p}}\, T^{(*)l}_{00}(q_\eta,\theta_\eta;\theta^*_{\pi
p},\phi^*_{\pi p}) \nonumber\\&=& -\int d\cos\theta_{\pi p}
^*\int_{q_\eta^{min}}^{q_\eta^{max}} q^2_\eta dq_\eta\,
\tau^{(*)l}_{00}(q_\eta,\theta_\eta;\theta^*_{\pi p},\phi^*_{\pi p}) \,,\label{Ic}
\\
I^s(\phi^*_{\pi p})\frac{d\sigma}{d\phi_\eta d\phi^*_{\pi p}}&=&
-\int d\cos\theta_{\pi p}^*\int_{q_\eta^{min}}^{q_\eta^{max}}
q^2_\eta
dq_\eta\,\frac{d\phi_\eta d\sigma_0}{d\phi^*_{\pi p}}\,
S^{(*)l}_{00}(q_\eta,\theta_\eta;\theta^*_{\pi p},\phi^*_{\pi p})
\nonumber\\&=&-\int d\cos\theta_{\pi p}^*\int_{q_\eta^{min}}^{q_\eta^{max}}
q^2_\eta dq_\eta\,
\sigma^{(*)l}_{00}(q_\eta,\theta_\eta;\theta^*_{\pi p},\phi^*_{\pi
  p})\,.\label{Is}
\end{eqnarray}
Using Eqs.\ (\ref{wima}), (\ref{taul}), and (\ref{sigmal}) one can easily show that
$I^c(\phi^*_{\pi p})$ and $I^s(\phi^*_{\pi p})$ are respectively even and odd functions
of the angle $\phi^*_{\pi p}$, i.e.
\begin{equation}\label{ICISsymmetry}
I^c(-\phi^*_{\pi p})=I^c(\phi^*_{\pi p})\,,\quad I^s(-\phi^*_{\pi p})=-I^s(\phi^*_{\pi
p})\,.
\end{equation}
Indeed, from the symmetry relation (\ref{wima})  with
  $\phi_\eta=0$ and thus $\phi_{pq}=\phi_{\pi p}$ and the definitions 
(\ref{taul}) and (\ref{sigmal})  follows 
\begin{eqnarray}\label{tausigmalsym}
\tau^{l}_{00}(q_\eta,\, \theta_\eta,\, \theta_{\pi p},\, -\phi_{\pi p})&=&
\tau^{l}_{00}(q_\eta,\, \theta_\eta,\, \theta_{\pi p},\, \phi_{\pi p})\,,\label{taulsym}
\\
\sigma^{l}_{00}(q_\eta,\, \theta_\eta,\, \theta_{\pi p},\, -\phi_{\pi p})&=&
-\sigma^{l}_{00}(q_\eta,\, \theta_\eta,\, \theta_{\pi p},\, \phi_{\pi
p})\,.\label{sigmalsym}
\end{eqnarray}
Furthermore from Eq.~(\ref{obs-star}) and the invariance in
Eq.~(\ref{Jacsym}) of the Jacobian one finds
\beqa
\tau^{(*)l}_{00} (q_\eta,\, \theta_\eta,\, \theta^*_{\pi p},\,
  -\phi^*_{\pi p})
&=&\tau^{l}_{00} (q_\eta,\, \theta_\eta,\, \theta_{\pi p},\,
  -\phi_{\pi p})\,|J(\cos\theta_{\pi p},-\phi_{\pi p};\cos\theta_{\pi p}^*,-\phi_{\pi
p}^*)|\nonumber\\
&=&\tau^{l}_{00} (q_\eta,\, \theta_\eta,\, \theta_{\pi p},\,
  \phi_{\pi p})\,|J(\cos\theta_{\pi p},\phi_{\pi p};\cos\theta_{\pi p}^*,\phi_{\pi
p}^*)|\nonumber\\
&=&\tau^{(*)l}_{00} (q_\eta,\, \theta_\eta,\, \theta^*_{\pi p},\,\phi^*_{\pi p})\,,\\
\sigma^{(*)l}_{00} (q_\eta,\, \theta_\eta,\, \theta^*_{\pi p},\,
  -\phi^*_{\pi p})
&=&\sigma^{l}_{00} (q_\eta,\, \theta_\eta,\, \theta_{\pi p},\,
  -\phi_{\pi p})\,|J(\cos\theta_{\pi p},-\phi_{\pi p};\cos\theta_{\pi p}^*,-\phi_{\pi
p}^*)|\nonumber\\
&=&-\sigma^{l}_{00} (q_\eta,\, \theta_\eta,\, \theta_{\pi p},\,
  \phi_{\pi p})\,|J(\cos\theta_{\pi p},\phi_{\pi p};\cos\theta_{\pi p}^*,\phi_{\pi
p}^*)|\nonumber\\
&=&-\sigma^{(*)l}_{00} (q_\eta,\, \theta_\eta,\, \theta^*_{\pi p},\,
  \phi^*_{\pi p})\,.
\eeqa
From these relations follow directly with the help of the definitions
in Eqs.~(\ref{Ic}) and (\ref{Is}) the noted symmetries of
Eq.~(\ref{ICISsymmetry}). 

\begin{figure}
\begin{center}
\includegraphics[scale=.8]{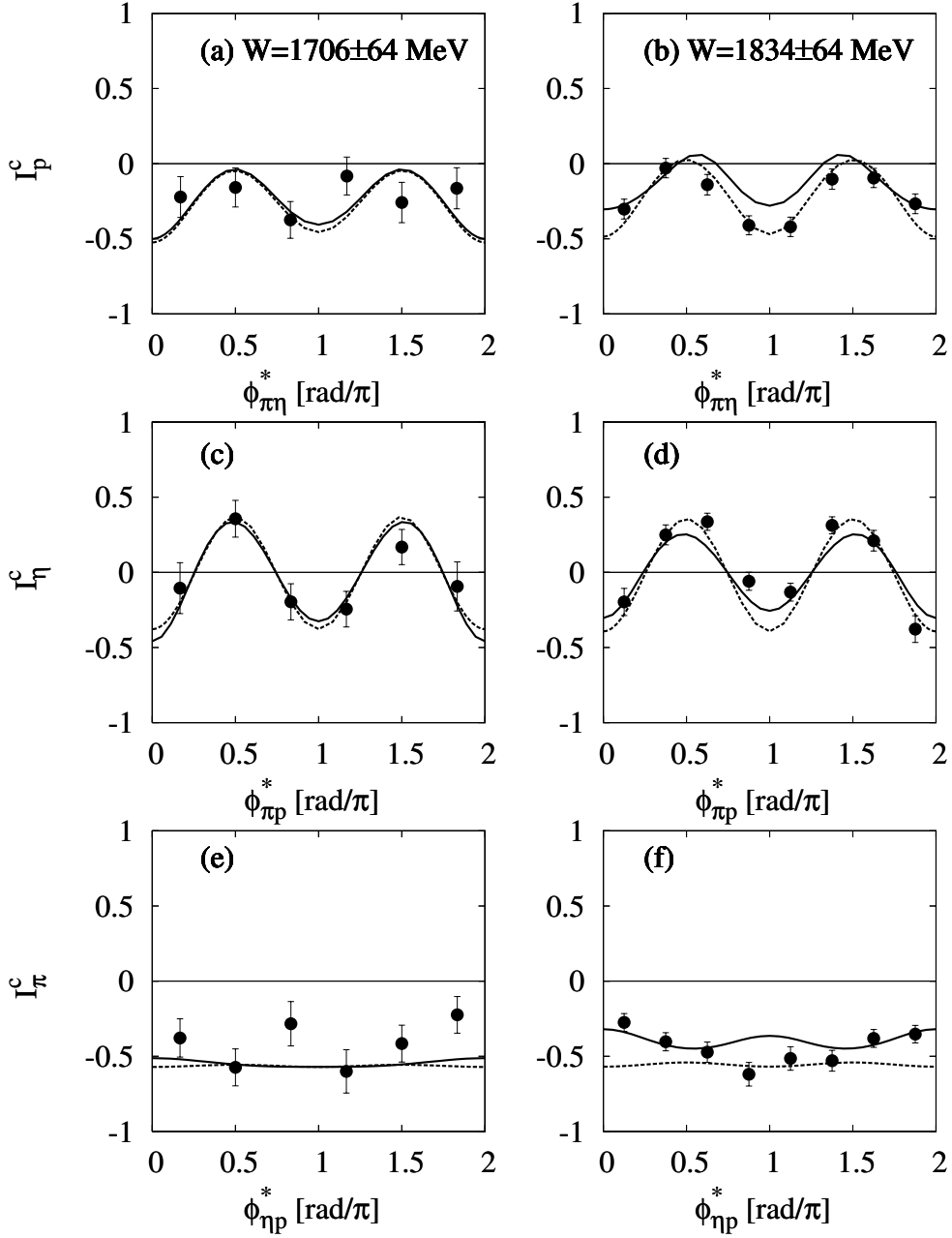}
\caption{The beam asymmetry $I^c$ calculated for two total c.m.\ energies $W$. The data
are from Ref.~\protect\cite{Gutz_Is} (only the statistical errors are shown). The
upper two panels (a) and (b) refer to an active proton, the two middle
panels (c) and (d) to an active eta, and the two lower panels (e) and
(f) to an active pion as function of the angle between the
corresponding reaction plane and decay planes as counted from the reaction plane (see
Fig.~\protect\ref{fig_kinematics_cm}). Notation of the curves as in
Fig.~\protect\ref{figFE}.} \label{figIc}
\end{center}
\end{figure}

\begin{figure}
\begin{center}
\includegraphics[scale=.8]{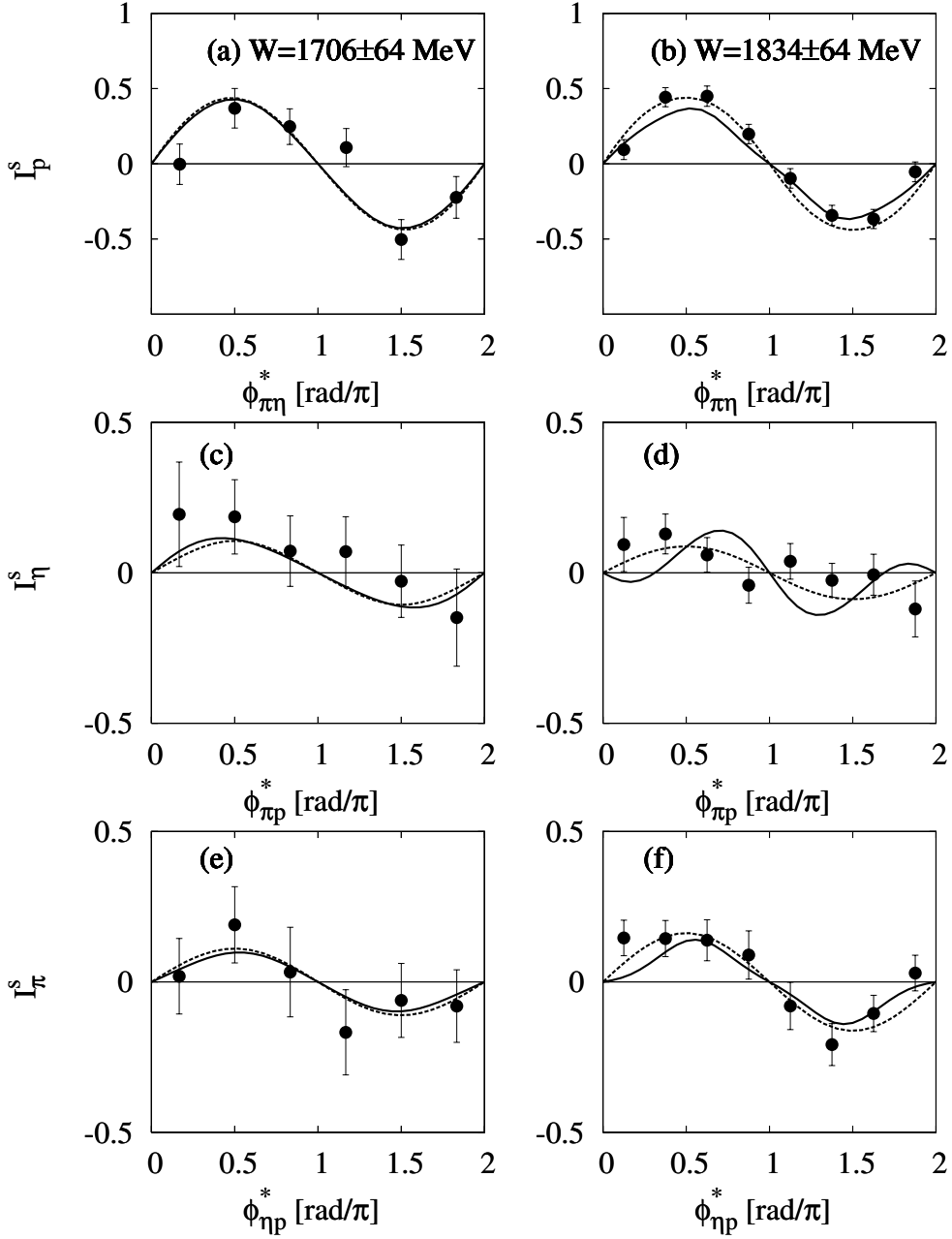}
\caption{Same as in Fig.~\protect\ref{figIc} for the beam asymmetry $I^s$.}
\label{figIs}
\end{center}
\end{figure}

In Figs.~\ref{figIc} and \ref{figIs} we compare our results with the data. In view of the
fact that the data were not included in the fit of the model parameters, the agreement is
reasonable. Already the single $D_{33}$ model (including only $D_{33}(1700)$ and
$D_{33}(1940)$) reproduces the experimentally observed shape and magnitude of the
observables, so that admixtures of other terms leads to relatively small corrections. Our
results are in general agreement with those obtained in Ref.\,\cite{Doring}, except, may
be, $I^s_\pi$ for which the model \cite{Doring} predicts vanishingly small values (see
Fig.~4 of the cited paper).

At the end of this section we will briefly return to circularly
polarized photons.  Without target polarization one has as
semi-exclusive cross section for the same experimental conditions as
above
\beq
\frac{d\sigma}{ d\phi_\eta d\phi^*_{\pi p}}=
\frac{d\sigma_0}{ d\phi_\eta  d\phi^*_{\pi p}}
\Big(1+P^\gamma_c\,T^{(*)c}_{00}\Big)\,,
\eeq
where only one beam asymmetry appears. In Ref.~\cite{Roberts} this
circular photon asymmetry was introduced with the notation $I^\odot$, i.e.\
\begin{eqnarray}\label{10}
\frac{d\sigma_0}{ d\phi_\eta  d\phi^*_{\pi p}} \,T^{(*)c}_{00}(\phi^*_{\pi p})&=&
\frac{d\sigma_0}{d\phi_\eta d\phi^*_{\pi p}}\,I^\odot(\phi^*_{\pi p})
=\frac12\frac{d\sigma^+-d\sigma^-}{d\phi_\eta d\phi^*_{\pi p}}\nonumber\\
&=&\int d\cos\theta_{\pi p}^*\int_{q_\eta^{min}}^{q_\eta^{max}}
q^2_\eta \,dq_\eta \,\tau^{(*)c}_{00}(q_\eta,\theta_\eta;\theta^*_{\pi p},\phi^*_{\pi p})\,,
\end{eqnarray}
where $d\sigma^{\pm}$ denotes the cross section corresponding to the photon beam
with a helicity $P_c^\gamma=\lambda_\gamma=\pm 1$,
respectively. Furthermore, like in Ref.~\cite{KashevAphi}
instead of $I^\odot$ we will consider an observable whose definition slightly differs
from Eq.~(\ref{10}), namely
\begin{equation} \label{10a}
W^c(\phi^*_{\pi p})=\frac{2\pi}{\sigma}\,
\frac{d\sigma}{d\phi_\eta  d\phi^*_{\pi p}}\, T^{(*)c}_{00}=
\frac{\pi}{\sigma}\frac{d\sigma^+-d\sigma^-}{d\phi_\eta  d\phi^*_{\pi p}}\,,
\end{equation}
with $\sigma$ being the unpolarized total cross section. According to the
definitions in Eqs.~(\ref{vPM}) and
(\ref{tau0c}), and the symmetry property in Eq.~(\ref{vminus}), $W^c$
is an odd function of the
argument $\phi^*_{\pi p}$ and therefore may be expanded into a sine-series
\begin{equation}\label{Wsin}
W^c(\phi^*_{\pi p})=\sum_nA_n\sin n\phi^*_{\pi p}\,.
\end{equation}
For further analyses it is convenient to have an analytic expression for $W^c(\phi^*)$
of Eq.~(\ref{10a}) and we neglect for simplicity the small background. Furthermore, as already
noted, in our energy region the reaction seems to be dominated by the $D_{33}$ wave
accompanied by relatively small admixtures of resonance states in other waves, in our
case $P_{33}$, $P_{31}$, and $F_{35}$. The latter contribute mainly as long as the
corresponding amplitudes can interfere with that coming from the $D_{33}$ excitation. In
this connection, we will retain in further relations only those terms which are linear
in the ``weak'' amplitudes. Then the integrand in Eq.~(\ref{35}), calculated up to the
first order in $t^{P_{31}}$, $t^{P_{33}}$, and $t^{F_{35}}$, reads
\begin{eqnarray}\label{40}
&&\tau^{(*)c}_{00}\simeq \left|\,t^{D_{33}}_{m_f\lambda}\right|^2+2\Re e\Big\{\,
t_{m_f\lambda}^{*\,D_{33}}t_{m_f\lambda}^{P_{31}}+
t_{m_f\lambda}^{*\,D_{33}}t_{m_f\lambda}^{P_{33}}+
t_{m_f\lambda}^{*\,D_{33}}t_{m_f\lambda}^{F_{35}}\Big\}-(\lambda\to -\lambda)\,.
\end{eqnarray}
Using Eqs.~(\ref{20}) through (\ref{35b}) in (\ref{40}) one obtains for the
asymmetry in Eq.~(\ref{10a})
\begin{equation}\label{45}
W^c(\phi^*_{\pi p})=A_1\sin\phi^*_{\pi p}+A_2\sin2\phi^*_{\pi p}\,,
\end{equation}
where the coefficients $A_1$ and $A_2$ are expressed in terms of resonance parameters and
are given in Appendix~\ref{appa4}. Of key importance is the fact that the first term in
(\ref{45}) is almost exclusively determined by the $D_{33}$ wave. The contributions of
other waves into $A_1$ are quadratic in the corresponding amplitudes and may therefore be
neglected. As a result, the ``weak'' resonances enter only into the second term of
Eq.~(\ref{45}) which is due to an interference of the amplitudes $t^{P_{31}}$,
$t^{P_{33}}$, and $t^{F_{35}}$ with the dominant $t^{D_{33}}$. In this respect, the $\sin
2\phi^*_{\pi p}$ admixture in the asymmetry $W^c(\phi^*_{\pi p})$ may be viewed as a signature of
positive parity states in $\pi^0\eta$ photoproduction.

\begin{figure}
\begin{center}
\includegraphics[scale=.75]{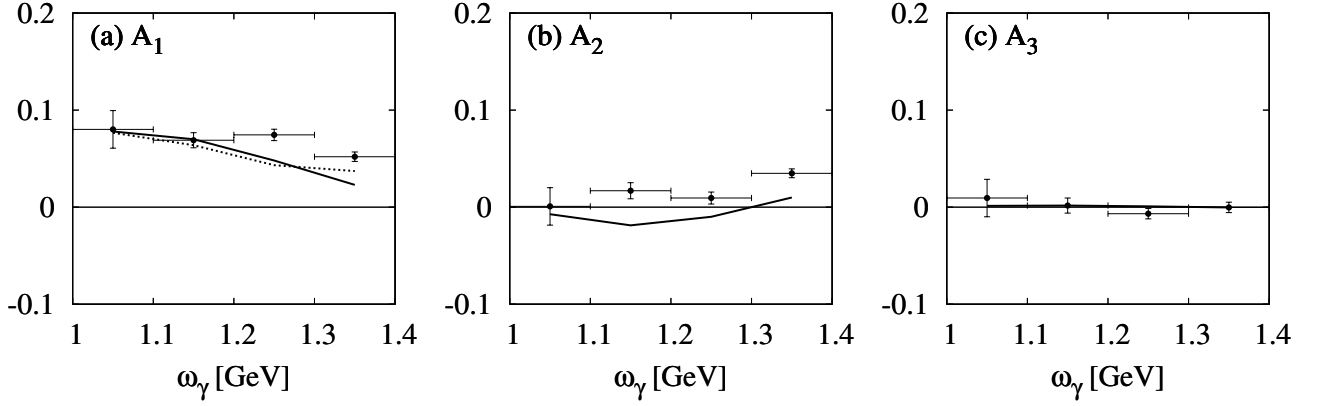}
\caption{Coefficients $A_n$ $(n=1,2,3)$ of the sine expansion (\protect\ref{Wsin}).
Notations as in Fig.~\protect\ref{figFE}. The data are from
Ref.~\protect\cite{KashevAphi}.} \label{fig3}
\end{center}
\end{figure}

In Fig.~\ref{fig3} we compare our calculation for $A_n$, $n=1,2,3$ with the results
obtained from the measurements of Ref.~\cite{KashevAphi}. As one can see, the single
$D_{33}$ reonance model reproduces rather well the coefficient $A_1$ in the whole energy interval.
As expected, addition of other resonances does not visibly change its value, since as
already noted the corresponding contributions are of second order in the ``small''
amplitudes. For $A_2$ the agreement is worse. In particular, the model
gives a wrong sign
of this coefficient. It is also worth noting that $A_2$ has a rather
small value at $\omega_\gamma\leq 1.3$\,GeV.
Unfortunately, the data do not allow us to find the reason of this fact, whether it
is a consequence of a general smallness of individual contributions,
or whether it is caused by an
accidental cancellation between different terms. The last coefficient $A_3$ is comparable
with zero, which is in line with our discussion above as well as with the model predictions.
In the general case, the term with $\sin 3\phi^*_{\pi p}$ would be due
to an interference of $D_{33}$
with negative parity resonances like $S_{31}$, $D_{35}$ {\it etc}. In this respect its
smallness may be considered as an indication of an insignificant role
of these states in this reaction.

\section{Conclusion}
In this work we have derived formal expressions for the differential
cross section and the recoil
polarization of $\pi\eta$ photoproduction on the nucleon including various polarization
asymmetries with respect to polarized photons and nucleons.

A general analysis allowing the determination of the moduli and relative phases of the four
independent photoproduction amplitudes requires a complete set of polarization experiments,
which for photoproduction of two pseudoscalar mesons is discussed, e.g., in
Ref.~\cite{Roberts}. However in the $\pi\eta$ case, due to the assumed dominance of the
$D_{33}$ wave, the information on bilinear combinations of the amplitudes may require
much less parameters. The situation is similar to that existing in $\eta$
photoproduction, which is known to be dominated by the $S_{11}$ wave in a wide energy
region. Making use of this fact has allowed, e.g., an almost model independent extraction of
the parameters of the resonance $D_{13}(1520)$ in a much cleaner way, than in $\pi$
photoproduction, where it overlaps with a multitude of other resonance states.

As noted above, according to the analyses of Refs.~\cite{Horn,FKLO}, in the energy
region below $\omega_\gamma=1.4$~GeV the main contribution beyond the
$D_{33}$ resonance should come
from the positive parity states $P_{33}$, $P_{31}$, and $F_{35}$ which reveal themselves
through their interference with the dominant $D_{33}$ amplitude. Our results show that
the corresponding ``small'' amplitudes may be identified, e.g., through their
contribution to the second Fourier coefficient $A_2$ in the sine-series for $W^c(\phi^*_{\pi p})$
in Eq.~(\ref{Wsin}).

It is also important to note, that the $D_{33}$ resonance decays
predominantly into an s-wave $\eta\Delta$ state.
As a result, in the single $D_{33}$ model (only the $D_{33}$ wave is included into the
amplitude) most of the polarization observables vanish. Therefore, the results of
polarization measurements are expected to be sensitive to even small admixtures of
``weak'' resonances.

A comprehensive program for single and double polarization measurements of the reaction
$\gamma p\to\pi^0\eta p$ is planned for the near future at MAMI and ELSA. The information
obtained by these new experiments will provide stringent constraints on the quantum numbers
of the resonance states entering the reaction amplitude.

\section*{Acknowledgment}
This work was supported by the Deutsche Forschungsgemeinschaft (SFB 443, SFB/TR16), and
the RF Federal program ``Kadry'' (contract P691). A.~Fix would like to thank the Institut
f\"ur Kernphysik of the Johannes Gutenberg-Universit\"at Mainz for the kind hospitality.

\appendix
\renewcommand{\theequation}{A\arabic{equation}}
\setcounter{equation}{0}

\section{Semi-exclusive differential cross section $\vec p\,(\vec\gamma,\eta)\pi p$ }
\label{appa2}

To derive the general expression for the semi-exclusive cross section we first introduce
the quantities
\begin{eqnarray} W_{IM}(q_\eta,\, \theta_\eta)&=& \int d\,\Omega_{\pi p}
\,w_{IM}^{1}(q_\eta,\, \theta_\eta,\, \theta_{\pi p},\, \phi_{pq}) \nonumber\\&=&
-\frac{\widehat I}{\sqrt{2}}\,\int d\,\Omega_{\pi p}\,c(q_\eta,\theta_\eta,\Omega_{\pi
p})\, \sum_{m_i m_i'}(-1)^{\frac{1}{2}-m_i} \left(
\begin{matrix}
\frac{1}{2}& \frac{1}{2}&I \cr m_i'&-m_i&M \cr
\end{matrix} \right)\nonumber\\
&& \sum_{m_f}t^*_{m_f 1 m_i'}(q_\eta,\, \theta_\eta,\, \theta_{\pi p},\, \phi_{pq}) \,t_{m_f -1
m_i}(q_\eta,\, \theta_\eta,\, \theta_{\pi p},\,
\phi_{pq})\,,\\
V_{IM}^\pm(q_\eta,\, \theta_\eta)&=&V_{IM}^1(q_\eta,\, \theta_\eta)\pm
V_{IM}^{-1}(q_\eta,\, \theta_\eta)\,,
\end{eqnarray} with
\begin{eqnarray}
V_{IM}^\mu(q_\eta,\, \theta_\eta)&=& \int d\,\Omega_{\pi p}\,v_{IM}^\mu(q_\eta,\,
\theta_\eta,\, \theta_{\pi p},\, \phi_{pq})\nonumber\\&=& \frac{\widehat
I}{\sqrt{2}}\,\int d\,\Omega_{\pi p}\,c(q_\eta,\theta_\eta,\Omega_{\pi p})\, \sum_{m_i
m_i'}(-1)^{ \frac{1}{2}-m_i} \left(
\begin{matrix}
 \frac{1}{2}& \frac{1}{2}&I \cr m_i'&-m_i&M \cr
\end{matrix} \right)\nonumber\\
&& \sum_{m_f}t^*_{m_f \mu m_i'}(q_\eta,\, \theta_\eta,\, \theta_{\pi p},\, \phi_{pq}) \,t_{m_f
\mu m_i}(q_\eta,\, \theta_\eta,\, \theta_{\pi p},\, \phi_{pq})\,.
\end{eqnarray}
Using now the property (\ref{vminus}), one finds with the help of
\begin{equation}
\int_0^{2\pi} d\phi_{\pi p} f(-\phi_{pq})=\int_0^{2\pi} d\phi_{\pi p} f(\phi_{pq})
\end{equation}
for a periodic function $f(\phi_{pq}+2\pi)=f(\phi_{pq})$ (please note
$\phi_{pq}=\phi_{\pi p}-\phi_\eta$), the relation
\begin{equation}
V_{IM}^{-1}(q_\eta,\, \theta_\eta)=\int d\,\Omega_{\pi p}\,v_{IM}^{-1}(q_\eta,\,
\theta_\eta,\, \theta_{\pi p},\,\phi_{pq})=(-1)^I\,\int d\,\Omega_{\pi
p}\,v_{IM}^1(q_\eta,\, \theta_\eta,\, \theta_{\pi p},\,
-\phi_{pq})^*=(-1)^I\,V_{IM}^1(q_\eta,\, \theta_\eta)^*\,,
\end{equation}
and thus
\begin{equation}
V_{IM}^\pm(q_\eta,\, \theta_\eta)=V_{IM}^1(q_\eta,\, \theta_\eta)\pm
(-1)^I\,V_{IM}^1(q_\eta,\, \theta_\eta)^* \,.\label{VIM}
\end{equation}
Correspondingly, using (\ref{wima}), one obtains
\begin{equation}
W_{IM}(q_\eta,\, \theta_\eta)^*=(-1)^I\,\int d\,\Omega_{\pi p}\,w_{IM}^1(q_\eta,\,
\theta_\eta,\, \theta_{\pi p},\,-\phi_{pq})=(-1)^I\,W_{IM}(q_\eta,\,
\theta_\eta)\,.\label{WIM}
\end{equation}
From the two foregoing equations we can conclude that $V_{IM}^+$ and
$W_{IM}$ are real for $I=0$  and imaginary for $I=1$, whereas $V_{IM}^-$ is imaginary for
$I=0$ and real for $I=1$. Therefore, according to (\ref{tau0c}) through (\ref{sigmal})
the following integrated asymmetries vanish
\begin{eqnarray} \int d\,\Omega_{\pi p}\,\tau_{IM}^\alpha&=&0\,\,\mbox{ for }\,\, \left\{\begin{array}{ll}
\alpha\in\{0,l\},&\,\,\mbox{and}\,\,I=1\\
\alpha\in\{c\},&\,\,\mbox{and}\,\,I=0\\
\end{array}\right\}\,,\\
\int d\,\Omega_{\pi p}\,\sigma_{IM}^\alpha&=&0\,\,\mbox{ for }\,\, \left\{\begin{array}{ll}
\alpha\in\{0,l\},&\,\,\mbox{and}\,\,I=0\\
\alpha\in\{c\},&\,\,\mbox{and}\,\,I=1\\
\end{array}\right\}\,.
\end{eqnarray}
Instead of using these results for deriving from~(\ref{diffcross}) the three-fold
semi-exclusive differential cross section, we prefer to start from the expression in
(\ref{diffcrossa}), and obtain
\begin{eqnarray}
\frac{d^3\sigma}{ dq_\eta
d\Omega_\eta}&=&
\sum_{I=0,1}P^p_I \,
\Big\{ \sum_{M=0}^I \frac{1}{1+\delta_{M0}}\,d^I_{M0}(\theta_s) \Re e \,[e^{iM\phi_{\eta
s}}\,(V_{IM}^+ +P^\gamma_c\,V_{IM}^-)] \nonumber\\
&&+\,P^\gamma_l\,\sum_{M=-I}^I
d^I_{M0}(\theta_s)\, \Re e \,[e^{i\psi_M}W_{IM}] \Big\}\,.\label{diffcross_incla}
\end{eqnarray}
This expression can be simplified using
the fact that $i^{\delta_{I1}}\,W_{IM}$, $i^{\delta_{I1}}\,V_{IM}^+$ and
$i^{1-\delta_{I1}}\,V_{IM}^-$ are real according to (\ref{VIM}) and (\ref{WIM}). The
latter two quantities can be written as
\begin{eqnarray}
i^{\delta_{I1}}\,V_{IM}^+&=&2\,\Re e\,(i^{\delta_{I1}}\,V_{IM}^1)\,,\\
i^{1-\delta_{I1}}\,V_{IM}^-&=&2\,\Re e\,(i^{1-\delta_{I1}}\,V_{IM}^1)=-2\,\Im
m\,(i^{-\delta_{I1}}\,V_{IM}^1)\,.
\end{eqnarray} Using now
\begin{eqnarray}
\Re e
\,[e^{iM\phi_{\eta s}}\,V_{IM}^+]&=& \Re e \,[e^{i(M\phi_{\eta
s}-\delta_{I1}\,\pi/2)}\,i^{\delta_{I1}}\,V_{IM}^+]=
2\,\Re e\,(i^{\delta_{I1}}\,V_{IM}^1)\,\cos[M\phi_{\eta s}-\delta_{I1}\,\pi/2]\,,\\
\Re e \,[e^{iM\phi_{\eta s}}\,V_{IM}^-]&=& \Re e \,[\frac{1}{i}\,e^{i(M\phi_{\eta
s}+\delta_{I1}\,\pi/2)}\, i^{1-\delta_{I1}}\,V_{IM}^-]= -2\,\Im
m\,(i^{-\delta_{I1}}\,V_{IM}^1)\,
\sin[M\phi_{\eta s}+\delta_{I1}\,\pi/2]\,,\\
\Re e \,[e^{i\psi_M}\,W_{IM}]&=& \Re e
\,[e^{i(\psi_M-\delta_{I1}\,\pi/2)}\,i^{\delta_{I1}}\,W_{IM}]=
i^{\delta_{I1}}\,W_{IM}\,\cos[\psi_M-\delta_{I1}\,\pi/2]\,,
\end{eqnarray}
we find as final form for the three-fold semi-exclusive differential cross section
(\ref{diffcrossc}).

\renewcommand{\theequation}{B\arabic{equation}}
\setcounter{equation}{0}
\section{The recoil polarization} \label{appa3}

For the recoil polarization, we have to evaluate according to (\ref{cartesian-P}) the
quantities $B_{M'}^\pm$ of (\ref{B_M}). From (\ref{trace-a}) we obtain for $M'=0,1$
\begin{eqnarray}
B_{M'}^\pm&=&\frac{(-1)^{M'}}{2(1+\delta_{M'0})} \sum_{I=0,1}P^p_I \, \sum_{M=-I}^I
\,e^{iM\phi_{\eta s}}\, d^I_{M0}(\theta_s)\,\Big[\widetilde v_{M';IM}^{1\,\pm}
+\widetilde v_{M';IM}^{-1\,\pm} \nonumber\\&&
+ P^\gamma_c\,(\widetilde v_{M';IM}^{1\,\pm} -\widetilde v_{M';IM}^{-1\,\pm} ) 
+P^\gamma_l\,(\widetilde w_{M';IM}^{1\,\pm} \,e^{-2i\phi_{\eta\gamma}} +\widetilde
w_{M';IM}^{-1\,\pm} \,e^{2i\phi_{\eta\gamma}}) \Big]\,,\label{trace-pol}
\end{eqnarray}
where for convenience we have defined
\begin{eqnarray}
\widetilde v/\widetilde w^{\mu;\,\pm}_{M';IM}&=& v/w_{1M';IM}^\mu \pm
v/w_{1-M';IM}^{\mu}\,.
\end{eqnarray}
One should note that
\begin{equation}
\widetilde v/\widetilde w^{\mu;\,+}_{0;IM}=2\,v/w_{10;IM}^\mu \quad\mbox{and}\quad
\widetilde v/\widetilde w^{\mu;\,-}_{0;IM}=0. \label{vtilde10}
\end{equation}
These quantities obey the obvious property
\begin{equation}
\widetilde v/\widetilde w^{\mu;\,\pm}_{-M';IM}=\pm\widetilde v/\widetilde
w^{\mu;\,\pm}_{M';IM}\,. \label{vwtilde}
\end{equation}

Then one obtains as final expression for the cartesian nucleon recoil polarization
components as defined in (\ref{cartesian-P}), including beam and target polarization
contributions,
\begin{eqnarray}
P_{x_i}\frac{d\sigma}{ d\vec q_\eta d\Omega_{\pi p}}= \sum_{I=0,1} &P^p_I & \Big\{ \sum_{M=0}^I
d^I_{M0}(\theta_s) \Big[\tau^{0}_{x_i;IM}\,\cos{(M\phi_{\eta s})}
+\sigma^{0}_{ x_i;IM}\,\sin{(M\phi_{\eta s})}\nonumber\\
&+&P^\gamma_c\,(\tau^{c}_{x_i;IM}\,\cos{(M\phi_{\eta s})}
+\sigma^{c}_{x_i;IM}\,\sin{(M\phi_{\eta s})})\Big]
\nonumber\\& +&P^\gamma_l\,\sum_{M=-I}^I d^I_{M0}(\theta_s)\,
\Big[\tau^{l}_{x_i;IM}\,\cos{\psi_M}+\sigma^{l}_{ x_i;IM}\,\sin{\psi_M}\Big]
\Big\}\,,\label{recoil-pol-p}
\end{eqnarray}
where the various beam, target and beam-target asymmetries are given by
\begin{eqnarray}
 \tau/\sigma^{0}_{ x/y; IM}&=&
\mp\frac{1}{\sqrt{2}(1+\delta_{M0})}\, \Re e/\Im m \,(\widetilde v_{1; IM}^{1;\,-}
+\widetilde v_{1;
  IM}^{-1;\,-} )
\,,\label{asy0xy}
\\
\tau/\sigma^{c}_{ x/y; IM}&=& \mp\frac{1} {\sqrt{2}}\Re e/\Im m  \,(\widetilde v_{1;
IM}^{1;\,-} -\widetilde v_{1; IM}^{-1;\,-})
\,,\label{asycxy}
\\
\tau/\sigma^{l}_{ x/y; IM}&=&\mp\frac{1} {\sqrt{2}} \Re e/\Im m  \,(\widetilde w_{1;
IM}^{1;\,-} ) \,.\label{asylxy}
\\
\tau/\sigma^{0}_{ z; IM}&=&\frac{1}{2(1+\delta_{M0})}\, \Re e/\Im m  \,(\widetilde v_{0;
IM}^{1;\,+} +\widetilde v_{0;
  IM}^{-1;\,+} )\nonumber\\
&=&\frac{1}{1+\delta_{M0}}\,\Re e/\Im m  \,(v_{10; IM}^{1}+v_{10; IM}^{-1})
\,,\label{asy0z}
\\
\tau/\sigma^{c}_{ z; IM}&=&\frac{1}{2} \Re e/\Im m  \,(\widetilde v_{0; IM}^{1;\,+}
-\widetilde v_{0;
  IM}^{-1;\,+})\nonumber\\
&=&\Re e/\Im m  \,(v_{10; IM}^{1}-v_{10; IM}^{-1})
\,,\label{asycz}
\\
\tau/\sigma^{l}_{ z; IM}&=&\frac{1}{2}
\Re e/\Im m  \,(\widetilde w_{0; IM}^{1;\,+} )\nonumber\\
&=&\Re e/\Im m  \,(w_{10; IM}^{1} ) \,,\label{asylz}
\end{eqnarray}
where we have used (\ref{vtilde10}) for $P_z$.

\renewcommand{\theequation}{C\arabic{equation}}
\setcounter{equation}{0}
\section{The expansion coefficients} \label{appa4}

The first two coefficients in the Fourier expansion of $W^c$ in Eq.~(\ref{Wsin}) may be derived
from the general expressions in Eqs.~ (\ref{20}), (\ref{35a}),  and
(\ref{35b}). Using the actual
resonance quantum numbers, one obtains after straightforward manipulations
\begin{eqnarray}\label{50}
A_1&=&\frac{\pi}{\sigma}\
\bigg[\left(A_{3/2}^{D_{33}}\right)^2+\frac13\left(A_{1/2}^{D_{33}}\right)^2\bigg]\int
\Im m\left(c^{*(1)}_{D_{33}}c^{(2)}_{D_{33}}\right)
\sin^2\theta_{\pi p}^*d\theta_{\pi p}^*\,d\omega_\eta,
\end{eqnarray}
\begin{equation}\label{50a}
A_2=-\frac{\pi}{\sigma}\ \bigg(F_{D_{33}P_{31}}+F_{D_{33}P_{33}}+F_{D_{33}F_{35}}\bigg).
\end{equation}
The individual terms on the r.h.s.\ of Eq.~(\ref{50a}) read
\begin{eqnarray}
\label{55a} F_{D_{33}P_{31}}&=&-\frac{4}{3\sqrt{3}}\ A_{1/2}^{D_{33}}A_{1/2}^{P_{31}}\int
\Im m\left(c_{D_{33}}^{*(1)}c_{P_{31}}^{(\eta
\Delta)}\right)\sin^2\theta^*_{\pi p} d\theta^*_{\pi p}\,d\omega_\eta\,,\\
\label{55b} F_{D_{33}P_{33}}&=&\frac{8}{3\sqrt{15}}\left(A_{3/2}^{D_{33}}A_{3/2}^{P_{33}}
-A_{1/2}^{D_{33}}A_{1/2}^{P_{33}}\right)\int \Im m\bigg[\
c_{D_{33}}^{*(1)}c_{P_{33}}^{(\eta\Delta)}
+\sqrt{\frac23}\bigg(c_{D_{33}}^{*(2)}c_{P_{33}}^{(\pi N^*)}p^2_{\pi p}\nonumber\\
&-&c_{D_{33}}^{*(1)}c_{P_{33}}^{(\pi N^*)}X_\pi q_\eta p_{\pi p}\bigg)
\bigg]\sin^2\theta^*_{\pi p} d\theta^*_{\pi p}\,d\omega_\eta,\quad \ \\
\label{55c}
F_{D_{33}F_{35}}&=&-\frac{1}{\sqrt{15}}\left(\sqrt{6}A_{3/2}^{D_{33}}A_{3/2}^{F_{35}}
+A_{1/2}^{D_{33}}A_{1/2}^{F_{35}}\right)\int \Im m\bigg[\
c_{D_{33}}^{*(1)}c_{F_{35}}^{(\eta\Delta)}-2\bigg(c_{D_{33}}^{*(2)}c_{F_{35}}^{(\pi
N^*)}p^2_{\pi p}\nonumber\\
&-&c_{D_{33}}^{*(1)}c_{F_{35}}^{(\pi N^*)}X_\pi q_\eta p_{\pi p}\bigg)
\bigg]\sin^2\theta^*_{\pi p} d\theta^*_{\pi p}\,d\omega_\eta\,,
\end{eqnarray}
where $X_\pi=m_\pi/(M_p+m_\pi)$.
In the expressions above, $\vec p_{\pi p}$ is as previously the relative $\pi p$ momentum. The
factors $c_R^{(\alpha)}$ $\alpha\in\{\eta\Delta,\pi N^*\}$ appear in the general ansatz
for the resonance amplitudes in Eq.~ (\ref{20}). For convenience we have introduced in
Eqs.~(\ref{55a})-(\ref{55c}) the following notations for the combinations of the
coefficients $c_{D_{33}}^{(\alpha)}$
\begin{equation}\label{57}
c_{D_{33}}^{(1)}=c_{D_{33}}^{(\eta\Delta)}+\frac{p}{q_\pi}\,c_{D_{33}}^{(\pi N^*)},\quad
\quad c_{D_{33}}^{(2)}=-\frac{q_\eta}{q_\pi}X_\pi\,c_{D_{33}}^{(\pi N^*)}.
\end{equation}

\renewcommand{\theequation}{D\arabic{equation}}
\setcounter{equation}{0}

\section{The $T$-matrix for an active proton} \label{appa5}
For an active proton, the partial wave decomposition of the
final state reads
\begin{eqnarray}
^{(-)}\langle \vec q_{\pi\eta}\,|&=&\frac{1}{\sqrt{4\pi}}
\sum_{l_{\pi\eta} m_{\pi\eta}}\widehat l_{\pi\eta}\,D^{l_{\pi\eta}}_{0,m_{\pi\eta}}( \phi_{\pi\eta},-\theta_{\pi\eta},-\phi_{\pi\eta})
^{(-)}\langle q_{\pi\eta} l_{\pi\eta} m_{\pi\eta}|\,,\\
^{(-)}\langle \vec p_p\, m_f|&=&\frac{1}{\sqrt{4\pi}} \sum_{l_p j_p m_p}\widehat l_p\, (l_p
0 \frac{1}{2} m_f|j_p m_f)\, D^{j_p}_{m_f,m_p}(\phi_p,-\theta_p,-\phi_p)
^{(-)}\langle p_p\, (l_p  \frac{1}{2})j_p m_p|\,,
\end{eqnarray}
where again $m_{\pi \eta}$ and $m_p$ refer to the photon momentum $\vec k$ as quantization
axis. Then we follow the same steps as in Eqs.~(\ref{mat1}) through (\ref{smallt}). With
the help of the multipole decomposition and the Wigner-Eckart theorem, one
obtains
\begin{eqnarray}
^{(-)}\langle q_{\pi\eta}\,l_{\pi\eta} m_{\pi\eta} ;p_p\, (l_p \frac{1}{2})j_p m_p|{\cal O}^{\mu L}_\mu|
\frac{1}{2} m_i\rangle &=&\sum_{J M_J}(-1)^{l_{\pi\eta}-j_p+J}\,\widehat{J}
\left(\begin{array}{ccc} l_{\pi\eta} & j_p & J \cr m_{\pi\eta} & m_p & -M_J\cr
\end{array}\right)
\left(\begin{array}{ccc}
J & L & \frac{1}{2} \cr -M_J & \mu & m_i\cr
\end{array}\right)\nonumber\\&&\hspace*{1.5cm}\times
\langle q_{\pi\eta}  \,p_p;(l_{\pi\eta}(l_ps)j_p)J||{\cal O}^{\mu L}|| \frac{1}{2}\rangle\,,
\end{eqnarray}
with the selection rule $m_p+m_{\pi\eta}=M_J=\mu+m_i$.
Rewriting the angular dependence
\begin{eqnarray}
D^{j_p}_{m_f,m_p}(\phi_p,-\theta_p,-\phi_p)\,
D^{l_{\pi\eta}}_{0,m_{\pi\eta}}(\phi_{\pi\eta},-\theta_{\pi\eta},-\phi_{\pi\eta})&=& d^{j_p}_{m_f,m_p}(-\theta_p)
\,d^{\,l_{\pi\eta}}_{0,m_{\pi\eta}}(-\theta_{\pi\eta})\, e^{i((m_p-m_f)\phi_p+m_{\pi\eta}\phi_{\pi\eta})}
\,,
\end{eqnarray}
and rearranging
\begin{eqnarray}
(m_p-m_f)\phi_p+m_{\pi\eta}\phi_{\pi\eta}&=&m_{\pi\eta}\phi_{pq}+(\mu+m_i-m_f)\phi_p
\end{eqnarray}
with $\phi_{pq}=\phi_{{\pi\eta}}-\phi_p$, one finds that the dependence on $\phi_p$ can be
separated, i.e.\
\begin{eqnarray}\label{small_tp}
T_{m_f \mu m_i}(\Omega_p,\Omega_{\pi\eta})&=& e^{i(\mu+m_i-m_f)\phi_p} t_{m_f \mu
m_i}(\theta_p,\,\theta_{\pi\eta},\,\phi_{pq})\,,
\end{eqnarray}
where the small $t$-matrix depends only on $\theta_p$, $\theta_{\pi\eta}$, and the relative
azimuthal angle $\phi_{pq}$.

The explicit form for the $t$-matrix in case of an active proton then reads
\begin{eqnarray}
t_{m_f \mu m_i}(\theta_{\pi \eta},\,\theta_p,\,\phi_{pq})&=& \frac{1}{2\,\sqrt{2\pi}} \sum_{L
l_{\pi \eta}m_{\pi \eta} l_p  j_p m_p J J M_J}i^L\,\widehat L\,\widehat J\,\widehat l_p \,\widehat j_p
\,\widehat l_{\pi \eta}\,(-1)^{J+l_{\pi \eta}-\frac{1}{2}+m_f-l_p -j_p}\nonumber\\
&&\times\left(\begin{array}{ccc} l_p &  \frac{1}{2} & j_p \cr 0 & m_f & -m_f\cr
\end{array}\right)
\left(\begin{array}{ccc} l_{\pi \eta} & j_p &  J \cr m_{\pi \eta} & m_p & -M_J\cr
\end{array}\right)
\left(\begin{array}{ccc}
J & L & \frac{1}{2} \cr -M_J & \mu & m_i\cr
\end{array}\right)\nonumber\\
&&\times\langle p_{\pi \eta} \,p_p; (l_{\pi\eta} (l_p \frac{1}{2})j_p)J||{\cal O}^{\mu L}||\frac{1}{2}\rangle
d^{l_{\pi\eta}}_{0,m_{\pi\eta}}(-\theta_{\pi\eta})\,d^{j_p}_{m_f,m_p}(-\theta_p)\,
e^{im_{\pi\eta}\phi_{p q}}\,.\label{smalltp}
\end{eqnarray}
Parity transformation leads to the
following property of the reduced matrix element
\begin{eqnarray}
(-1)^{l_{\pi \eta}+l_p+L}\langle p_{\pi \eta} \, p_p; ( l_{\pi\eta}
(l_{\pi p}\frac{1}{2})j_{\pi p})J|| {\cal O}^{-\mu 
L}||\frac{1}{2}\rangle&=& \langle p_{\pi p}\, p_p; (l_{\pi\eta}
(l_{\pi p} \frac{1}{2})j_{\pi p} )J||{\cal O}^{\mu L}|| 
\frac{1}{2}\rangle\,,
\end{eqnarray}
which in turn gives the symmetry property of Eq.~(\ref{symmetry}).

\end{document}